\documentclass[aps,amsfonts,amsmath,prd,preprint,nofootinbib]{revtex4}
\pdfoutput=1
\newcommand{\beq}{\begin{equation}}
\newcommand{\eeq}{\end{equation}}
\usepackage{graphicx}
\usepackage{natbib}

\usepackage{xcolor}

\usepackage{amsmath}

\newcommand{\be}{\begin{equation}}
\newcommand{\ee}{\end{equation}}
\newcommand{\bea}{\begin{eqnarray}}
\newcommand{\eea}{\end{eqnarray}}
\newcommand{\bi}{\begin{itemize}}
\newcommand{\ei}{\end{itemize}}
\newcommand{\pa}{\partial}

\begin{document}

\title{Internal Excitations of Global Vortices}

\author{Jose J. Blanco-Pillado{$^{1,2,}$}\footnote{josejuan.blanco@ehu.eus},
Daniel Jim\'enez-Aguilar{$^{1,}$}\footnote{daniel.jimenez@ehu.eus},
Jose M. Queiruga{$^{1,}$}\footnote{josemanuel.fernandezq@ehu.eus},
and Jon Urrestilla{$^{1,}$} \footnote{jon.urrestilla@ehu.eus}}

\affiliation{ $^1$ Department of Physics, UPV/EHU, 48080, Bilbao, Spain \\
$^2$ IKERBASQUE, Basque Foundation for Science, 48011, Bilbao, Spain \\
}

\begin{abstract}
We investigate the spectrum of linearized excitations of global vortices in $2+1$ dimensions.
After identifying the existence of localized excitation modes, we compute the decay
time scale of the first two and compare the results to the numerical evolution of the full non-linear
equations. We show numerically how the interaction of vortices with an external source
of radiation or other vortices can excite these modes dynamically. We then simulate 
the formation of vortices in a phase transition and their interaction with a thermal 
bath estimating the amplitudes of these modes in each case. These numerical experiments
indicate that even though, in principle, vortices are capable of storing a large amount 
of energy in these internal excitations, this does not seem to happen dynamically.
We then explore the evolution of a  {\it network of vortices} in an expanding (2+1) dimensional
background, in particular in a radiation dominated universe. We find that vortices are 
still excited after the course of the cosmological evolution but again the level of 
excitation is very small. The extra energy in the vortices in these cosmological 
simulations never exceeds the $1\%$ level of the total mass of the core of the vortex.
\end{abstract}

\maketitle

\section{Introduction}
 
 Topological defects are solitonic field theory solutions that appear as part of the spectrum in many high energy theories beyond the 
 Standard Model \cite{Vilenkin:2000jqa}. These models predict the existence of a sequence of symmetry breaking phase transitions in the
early universe where these objects can be created \cite{Kibble:1976sj}. Depending on the pattern of symmetry breaking one can form monopoles, strings, 
domain walls or a combination of them if more than one transition occurs. Their topological stability allows them to survive for long periods of
time leaving their imprint in some of the cosmological observables that we can probe today.

In this paper we will be interested in global vortices as a first step towards understanding the dynamics of their $3+1$ dimensional counterparts, global strings. 
These cosmic strings are formed whenever a $U(1)$ global symmetry is spontaneously broken. As we will describe later on in detail these strings are coupled
to the long range field, the massless Goldstone mode \cite{Vilenkin:1982ks}. After their formation, the network of strings will
evolve losing energy by the emission of these massless modes, the would be axion particles. One of the most interesting aspects of these studies is the
idea of associating the axion to the dark matter in the universe \cite{Preskill:1982cy,Abbott:1982af,Dine:1982ah}. It is therefore clear that in order to give an accurate
calculation of the density of axions one needs to understand the evolution of these networks. This program is currently being
pursued by several different groups that have used lattice field theory simulations to infer the long term behaviour of
these networks \cite{Kawasaki:2018bzv,Vaquero:2018tib,Gorghetto:2018myk,Gorghetto:2020qws,Fleury:2015aca,Fleury:2016xrz,Klaer:2017qhr,Hindmarsh:2019csc,Hindmarsh:2021vih} (For earlier studies see \cite{Sikivie:1982qv,Hagmann:1990mj,Hagmann:2000ja,Yamaguchi:1998gx,Yamaguchi:2002sh}).
Some of the properties of these networks are still under debate. In particular there is currently
some controversy about the asymptotic nature of the scaling limit on these networks or the spectrum
of axions being produced by the network.

One of the key issues in this debate is whether one is able to have a sufficiently large dynamic range in these
simulations to be able to extrapolate the results from the small simulation to a cosmological setting. This is
particularly important for the proper evolution of string loops since they will in turn dominate the axion spectrum. It is therefore
crucial to understand the underlying physics that controls the dynamics of these axionic strings. The effective
theory for the string dynamics in these models is complicated by their coupling to low energy degrees of freedom
propagating outside of the string, the axions. This means that the Nambu-Goto action for relativistic strings should 
be supplemented in these models by a Kalb-Ramond term \cite{Kalb:1974yc}. This coupling is important not only to
understand the string radiation but also its subsequent motion \cite{Vilenkin:1986ku,Garfinkle:1988yi,Dabholkar:1989ju}. In this approach 
one disregards the dynamics of the massive (radial) mode for the complex scalar field assuming that the string is not significantly curved to 
excite these modes.

These ideas have been investigated numerically. Field theory simulations of global strings have been done 
for oscillating configurations of small amplitude. The results seem to be in pretty good agreement with the previous effective 
action \cite{Drew:2019mzc} where a comparison with analytic estimates for backreaction has been taken into
account \cite{Battye:1993jv,Battye:1995hw,Battye:1998mj}. However, recent simulations of 
global string loops created from the intersection of long strings in a box suggest a different picture where 
a combination of massless and massive radiation has been observed \cite{Saurabh:2020pqe}. This represents an important departure
from the previous effective theory for a global string that does not consider any massive radiation. It has been suggested that this could
be due to the presence of internal excitations of strings that lead to a transfer of energy from the string
motion to massive states leaving the string. 

These results remind us the situation with local cosmic strings. In that case, field theory simulations of
individual string configurations initially prepared from exact solutions have shown that local cosmic strings do indeed follow 
the Nambu-Goto dynamics \cite{Olum:1998ag,Olum:1999sg}. However, the results obtained from 
simulations of string networks seem to describe a different behaviour \cite{Hindmarsh:2017qff}. It has also been suggested that
internal degrees of freedom found in these models \cite{Arodz:1991ws} can play a significant role in that 
case \cite{Hindmarsh:2017qff,Hindmarsh:2021mnl}\footnote{These excitation modes in local strings have also been discussed in
\cite{Goodband:1995rt,Kojo:2007bk,Alonso-Izquierdo:2015tta,Alonso-Izquierdo:2016pcq}.}. Note, however, that this effect was not observed
in \cite{Matsunami:2019fss} where these internal modes do not seem to be clearly excited. 

 The existence of these internal bound states is a generic feature of most field theory models with solitons. Furthermore, 
 their excitation could lead to a variation of the expected dynamics of these solitons. It is therefore important to understand how these
modes are excited and how they interact with other degrees of freedom as well as their possible
decay. We have recently explored these issues in the simplest soliton model; the $1+1$ kink soliton
solution in the $\phi^4$ theory \cite{Blanco-Pillado:2020smt}.  In this paper we will continue our exploration of these
bound states on solitons by studying the dynamics of these internal excitations in global string vortices in $(2+1)-$dimensions. 
We will also comment on how these modes can also exist on a $3+1$ dimensional string but we leave the exploration 
of the relevance of these modes for realistic cosmological models for a future publication.

The organization of the paper is the following. In Section (II) we introduce the field theory model for the
global (axionic) strings and describe the static configuration for the cross section of the string. This cross section
describes the global vortex we will be discussing in the rest of the paper. In Section (III) we 
study the linearized equations of motion for perturbations around the static vortex solution presented
earlier and identify the bound state solutions of those equations. We investigate analytically the decay rate of these 
bound states in Section (IV). In Section (V) we compared these analytic calculations to the results obtained 
using lattice field theory simulations. We perform a series of 
numerical experiments to probe the possible excitation of the bound states by an external source of
radiation in Section (VI). In Section (VII) we explore the amplitude of these internal excitations in the formation
of vortices in a $(2+1)$-dimensional phase transition as well as the amplification of these modes by their
interaction with a thermal bath. In Section (VIII) we perform a series of cosmological simulations of the evolution
of a collection of vortices in an expanding $2+1$ dimensional universe and obtain the level of excitation
of these vortices as they undergo this cosmological evolution. Finally we conclude in Section (IX) with a brief discussion of the results and
their impact to the evolution of axionic strings in numerical simulations as well as in a realistic
cosmological setting.\\

Some examples of the simulations we have performed in this paper
can be found at:  \url{http://tp.lc.ehu.es/earlyuniverse/global-vortex-simulations/}.

\section{Global string background}
\label{section-string-background}

The simplest field theory model that allows for smooth global string configurations is of the form
\beq 
\label{themodel}
\mathcal{L}=\pa_\mu\phi^\star\pa^\mu \phi-\frac{\lambda}{4}\left(\phi^\star\phi-\eta^2\right)^2\,,
\eeq
where $\eta$ is the only energy scale in the problem and we have denoted by 
$\lambda$ the quartic self-coupling of the field $\phi$. This model is clearly invariant under a global $U(1)$ symmetry; however, the potential
leads to the spontaneous symmetry breaking where the vacuum manifold of this theory is parametrized by  $|\phi|^2 = \eta^2$.
Perturbations around this type of configuration describe the two degrees of freedom present
in these vacua; the field associated with the radial excitations whose mass is $m_r = \sqrt{ \lambda} \eta$
and the massless Goldstone boson that parametrizes perturbations on the field's phase.

\begin{figure}[h!]
\includegraphics[width=11cm]{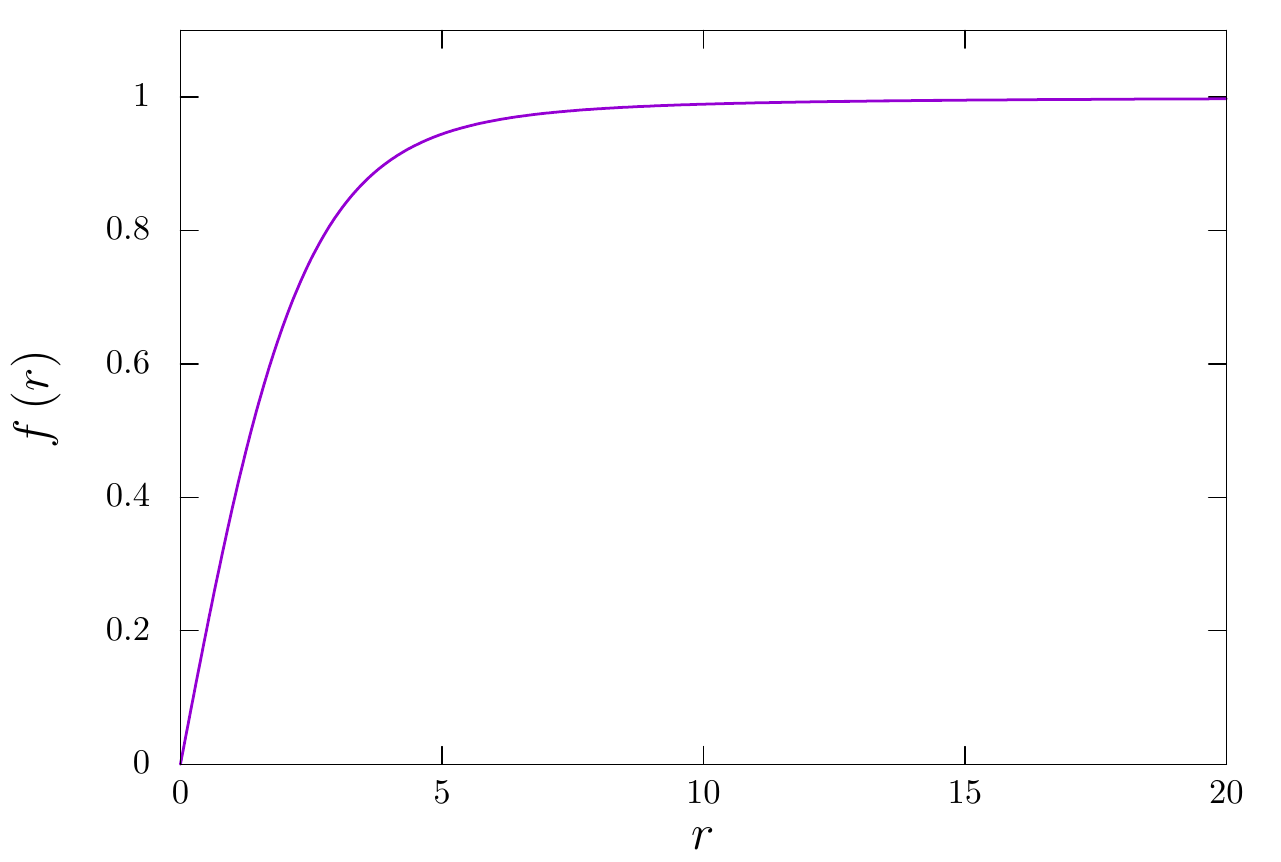}
\caption{Profile function, $f\left(r\right)$, for a static vortex with unit winding.}
\label{fig:f r}
\end{figure}

The equations of motion arising from this Lagrangian can be found to be
\beq
\label{eom}
\partial_{\mu}\partial^{\mu} \phi + \frac{\lambda}{2}\left(\vert \phi \vert^2-\eta^2\right)\phi=0~.
\eeq
We are interested in string like configurations in this model, so we will consider a
 $z-$independent ansatz for the field of the form
 \beq
 \phi_v({\bf x}) =\eta f(\rho)e^{iN\theta}~,
 \eeq
 where $\rho$ and $\theta$ denote the usual polar coordinates on the plane perpendicular to the 
 string and $N$ is the winding number of the 2 dimensional vortex solution we seek. After rescaling
the radial coordinate by defining $r=\sqrt{\lambda \eta^2}\rho = m_r \rho$, the field equation for the profile function $f(r)$ in this ansatz is given by
\beq\label{global-field-eq}
\frac{d^2 f}{dr^2}+\frac{1}{r}\frac{df}{dr}-\frac{N^2}{r^2} f-\frac{1}{2}\left(f^2-1\right)f=0,
\eeq
where we will take the appropriate boundary conditions with $f(0)=0$ at the center of 
our coordinates, so that the solution remains smooth there; and $f(\infty)=1$, so that the
solution approaches the vacuum asymptotically.  Using the equation of motion one can show that the approximate 
behaviour of $f(r)$ for $r\rightarrow \infty$ is
$f(r)=1-N^2/r^2+\mathcal{O}(1/r^4)$, while for $r\rightarrow 0$, $f(r)= c_n r^N+ ... $. 
We show in Fig. {\ref{fig:f r}} the profile for the case with $N=1$.

The energy per unit of length of this string can be written as
\beq\label{global-ener}
E=2\pi  \eta^2 \int_0^\infty r dr\left(\left(\frac{df}{dr}\right)^2+\frac{N^2}{r^2}f^2+\frac{1}{4}\left(f^2-1\right)^2\right)~.
\eeq
Plugging the profile for the function $f(r)$ in this expression one can see two different contributions
to the energy density. One that is concentrated at the center of the vortex, the region $r < \delta \approx {\cal O}(1)$, associated with the
massive degree of freedom, and another one that comes from the variation of the phase mode around the string. The
first one gives a finite contribution to the energy per unit length while the second one leads to a logarithmic
divergence. 

 In a realistic context this divergence is cured by the presence of another string at some distance, $r=R$. This 
 cuts off the divergent contribution so we finally arrive to an expression of the energy for an $N=1$ vortex of the form
 \beq
E_0  \approx E_{\text{core}}+ 2\pi \eta^2 \log \left( {R\over \delta} \right)~.
 \eeq

Using the profile for the function $f(r)$ and performing the integrals described in Eq. (\ref{global-ener}),
we can estimate the values of the two different parts of the energy for the global vortex as
\beq
E_0 \approx 4.9 \eta^2 + 2\pi \eta^2 \log \left({ {\sqrt{\lambda \eta^2}\rho_{\text{max}}}\over 2.15} \right)=4.9 \eta^2 + 2\pi \eta^2 \log \left({ r_{\text{max}}}\over 2.15 \right).
\eeq
This expression gives a good fit for the total energy of the vortex as a function of the distance from the center $\rho_{\text{max}}$.

The existence of these two different contributions to the energy per unit length for global strings is also
reflected on the description of their dynamics. The Nambu-Goto effective action for relativistic strings is,
in this case, supplemented by the existence of a term that describes the coupling of the string to the Goldstone
mode. It turns out that there is a somewhat simpler description of this coupling in terms of a 2-form
potential $B_{\mu \nu}$ that is dual to the Goldstone mode in $4$ dimensions \cite{Davis:1988rw}. This field is naturally
coupled to the string worldsheet through the so-called Kalb-Ramond action \cite{Vilenkin:1986ku, Davis:1988rw}. This effective
action can therefore be used to compute the radiative decay of these strings into the massless mode.

In the following sections we will study the motion of several global string vortices, the $2+1$ dimensional
version of this theory. In this case the vortices are point-like objects in $2$ spatial dimensions and their
coupling to the Goldstone boson can be described by assigning these vortices an electric charge. Their
motion and radiation can then be studied in this limit where the massive radiation is neglected. We can then
compare this simplified dual description with the full numerical evolution in terms of the complex scalar field.

\section{Spectrum of perturbations}

In this section, we would like to identify the spectrum of perturbations around the background solution for the
global vortex discussed earlier. As we show in detail in Appendix \ref{appendix-perturbations},
this spectra is composed of discrete bound states, translational zero modes as well as a continuum of scattering states of the corresponding Schr\"odinger-like
operators for the linearized equations of motion of the fluctuations. Here we will concentrate on the properties 
of the bound states since they will be the main focus of our paper\footnote{You can find the details about other possible
states in  Appendix  \ref{appendix-perturbations}.}.

In particular, let us consider perturbations of the form
\beq
\label{pertur-1}
\phi({\bf x},t)= \phi_v({\bf x})+ A ~s(r) e^{i \theta} \cos\left(\omega t\right)~,
\eeq
where $s(r)$ is a real function that depends only on the radial coordinate
$r$ and $\phi_v({\bf x}) = f(r) e^{i \theta}$ denotes the dimensionless background
solution for the $N=1$ vortex where $f(r)$ satisfies Eq. (\ref{global-field-eq}). The
field configuration in Eq. (\ref{pertur-1}) will then obey the equations of motion given in Eq. (\ref{eom})
at a linear order if $s(r)$ satisfies,
\beq
\label{eigen-re}
-\nabla_r^2 s +\frac{1}{r^2}s+\frac{1}{2}\left(3f^2-1\right)s=\omega^2s ~.
\eeq

This is a Schr\"odinger type equation for the function $s(r)$ where the combination
$U(r)=\frac{1}{2}\left(3f^2-1\right)+ \frac{1}{r^2}$ plays the role of the potential. The asymptotic form of this
potential, where $f(r) \rightarrow 1$, indicates the presence of a continuous spectrum
of perturbations starting at $\omega_c =1$. In this limit, these solutions
correspond to the massive excitations of the radial component of the field where we
have imposed cylindrical symmetry.

We can also look for solutions of this equation in the form of normalized bound states
whose frequencies are below $\omega_c$. It can be shown that, due to the
slow decay of the potential at large distances from the core, there are infinitely many bound modes 
below $\omega_c$ \footnote{ We thank T. Romańczukiewicz for pointing this out to us.}.
 (See p. 117 of \cite{Landau1981Quantum} for a description of this type of behaviour in Quantum Mechanics.). A numerical scan of the possible
frequencies has uncovered the existence of the first two such bound states with frequencies
\bea
\omega_1^2 = 0.8133,\, \\
\omega_2^2 = 0.9979~.
\eea
These frequencies are given in units of $\lambda \eta^2$, in other words, in 
units of the mass of the radial mode in the vacuum \footnote{The first bound state had been already identified
in  \cite{Goodband:1995rt} and more recently in  \cite{Saurabh:2020pqe} but the higher ones appear to have been missed by
these earlier papers.}. Higher frequency modes belonging to the infinite series described earlier are all packed very close to the threshold of scattering 
states\footnote{One can use the asymptotic behaviour of the effective potential in our case,
namely, $U(r) \approx 1 - \frac{2}{r^2}$ to estimate the energy separation for these modes. See for example \cite{MF}, vol II, page 1665.}. More importantly, 
they are highly delocalized with respect to the vortex core, hence, in order to study them numerically 
one would require very large simulation boxes. Therefore, we will restrict the subsequent numerical analysis to the first
two bound modes, although the basic properties obtained for them may be easily extended to the other modes.

\begin{figure}[h!]
\includegraphics[width=8cm]{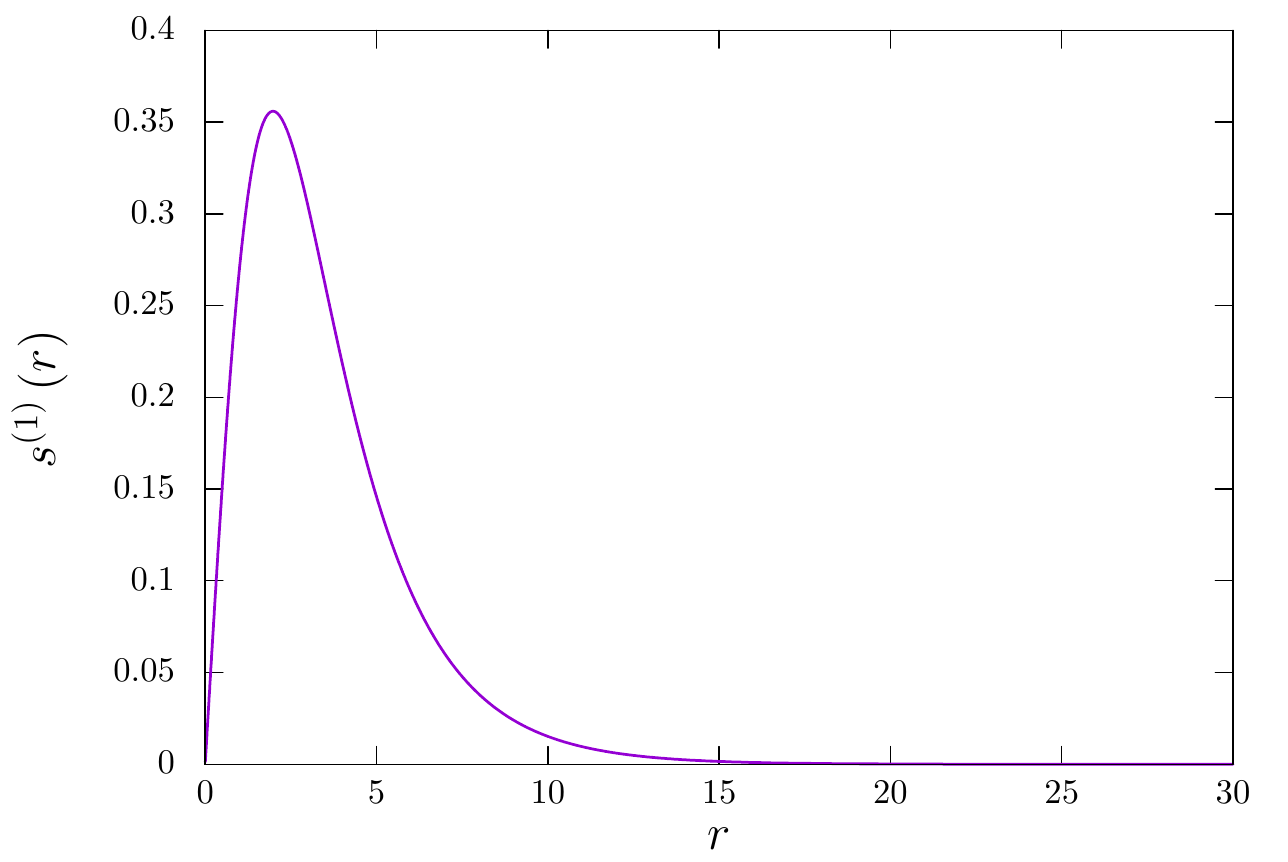}\includegraphics[width=8cm]{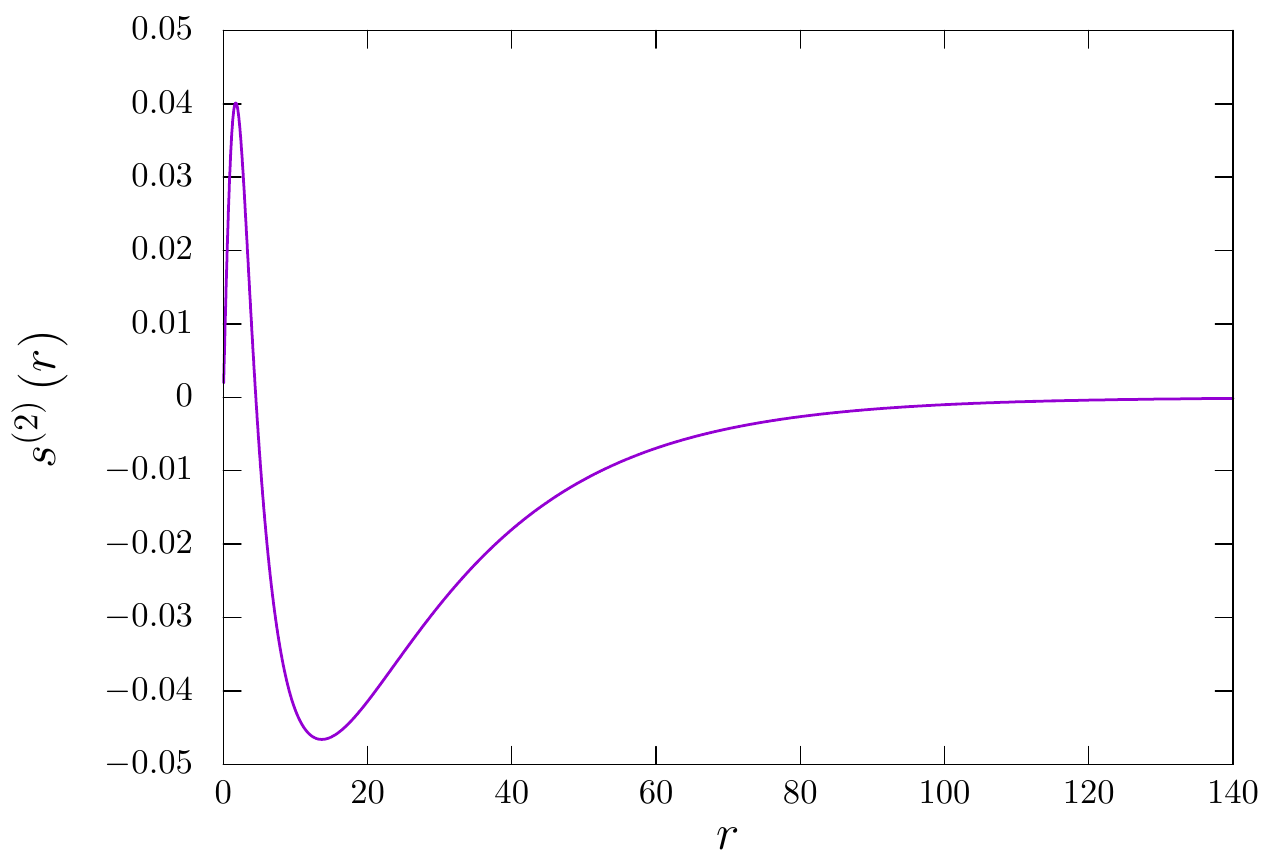}
\caption{Normalized bound state mode functions:  the first one $s^{(1)}\left(r\right)$ on the left,  and the second one $s^{(2)}\left(r\right)$ on the right. }
\label{fig:first bound}
\end{figure}

We show in Fig.~\ref{fig:first bound}  the profile for the first two mode functions. Note that there is
quite a big difference between these two bound states, not only in their extent but also in their
shape. The first state represents a fairly localized mode with support mostly over the core of the 
vortex. Furthermore, since it is a solution of the linearized equations of motion for the perturbations, 
one may think that this localized state will oscillate forever without dissipation. This suggests an interpretation
of this mode as a massive particle trapped on the core of the vortex. On the other hand, the effect of exciting this mode 
creates an oscillation of the apparent width of the vortex. In many respects, this mode is very similar 
to the \textit{shape mode} found in the $1+1$ dimensional kink solution \cite{Blanco-Pillado:2020smt}.

The situation for the second mode is somewhat similar, except that it is not so well localized. Its mode 
function spreads over a much larger region than the soliton core. This property can be
traced back to the fact that the frequency eigenvalue for this mode is very close to the continuum
so one can say that it is much more loosely trapped by the vortex. Moreover, its mode
function has a node at some distance from the center of the string as expected from a higher excitation mode.
This also means that its effect on the vortex width is less straightforward to interpret. Similar conclusions can be extended for modes above the second. We will collectively 
denote these modes as ``internal excitations" of the vortex. Finally, as we discuss in Appendix \ref{appendix-perturbations},
we have not found any other bound state in this model without cylindrical symmetry apart from the zero modes.

\section{Decay rate for the perturbations}

In the previous section we identified two excitations of the vortex that correspond to localized perturbations
of the radial degree of freedom of the complex scalar field. The calculation of the mode functions at the linear
level shows that their frequencies are not large enough to propagate in the vacuum. This would suggest that
the perturbations will live indefinitely. However, this argument is based on the linear theory. Non-linear interactions 
present in this model should allow for their slow decay. This is basically possible since higher order terms in the Lagrangian 
obtained from this oscillating perturbation lead to contributions that can act as sources with a higher frequency for the radiation
modes. These higher frequencies are above the continuum threshold and can propagate outside the vortex.
We therefore expect that, taking into account non-linearities, an initially small amplitude perturbation will decay
but with quite a long lifetime. This type of argument has been used in the past to compute the decay
rate of excitations in the $1+1$ dimensional kink soliton \cite{Manton:1996ex}. This analytical computation has been shown to 
agree with numerical calculations recently performed in \cite{Blanco-Pillado:2020smt}. Here we will follow a similar type of reasoning
and compute the expected time scale for both internal excitations of the vortex.

\subsection{The first localized mode}

We start by considering the general time dependent profile for the vortex excited by the presence of the first
localized mode, namely a field of the form
\beq
\phi({\bf x},t) = f(r) e^{i \theta} + A_1(t) s^{(1)}(r)  e^{i \theta} + \left[ \eta_1(r,t) + i~\eta_2(r,t) \right] e^{i \theta}\,,
\eeq
where the $\eta_{j}(r,t)$ terms represent the scattering modes that will contribute to the radiation field \footnote{Where the form
of the scattering states have been taken to have the same cylindrical symmetry as the source of the perturbations.}. Substituting
this expression into the equations of motion we obtain at $\mathcal{O}(A_1^2(t))$ order
\bea
&&\pa_t^2 \eta_1(r,t)-\pa_r^2  \eta_1(r,t)-\frac{\pa_r  \eta_1(r,t)}{r}+\frac{\eta_1(r,t)}{r^2}+\frac{1}{2}\left(3 f^2(r)-1\right)\eta_1(r,t)+\label{re-rad}\\
&&+s^{(1)}(r)\left(\frac{d^2 A_1(t)}{dt^2}+\omega_1^2 A_1(t)\right)+\frac{3}{2} A_1^2(t)\left(s^{(1)}(r)\right)^2f(r)=0\nonumber\,,\\ \nonumber\\\nonumber\\\
&&\pa_t^2 \eta_2(r,t)-\pa_r^2  \eta_2(r,t)-\frac{\pa_r  \eta_2(r,t)}{r}+\frac{\eta_2(r,t)}{r^2}+\frac{1}{2}\left( f^2(r)-1\right)\eta_2(r,t)=0\,,\label{im-rad}
\eea
where we have used (\ref{global-field-eq}) and (\ref{eigen-re}). The equation for $\eta_2(r,t)$ is not sourced by
the excitation of the first localized mode, so it corresponds to the usual equation for the scattering modes modified by the presence of the vortex.
We  can multiply Eq. (\ref{re-rad}) by $s^{(1)}(r)$ and noticing that $s^{(1)}(r)$ is orthogonal to $\eta_1(r)$, in the sense that the $2d$ integral
of these modes is zero, we obtain the approximate expression
\be
\frac{d^2 A_1(t)}{dt^2}+\omega_1^2 A_1(t)+\frac{3}{2} A_1^2(t)\alpha = 0,
\ee
where
\be
\alpha=\int dr r \left(s^{(1)}(r)\right)^3f(r).
\ee
Using this equation for $A_1(t)$  in (\ref{re-rad}) we finally obtain
\bea\label{eigen-rad-re}
&&\pa_t^2 \eta_1(r,t)-\pa_r^2  \eta_1(r,t)-\frac{\pa_r  \eta_1(r,t)}{r}+\frac{\eta_1(r,t)}{r^2}+\frac{1}{2}\left(3 f^2(r)-1\right)\eta_1(r,t)=\label{re-rad2}\\
&&\frac{3}{2}A_1^2(t)\left(\alpha s^{(1)}-\left(s^{(1)}(r)\right)^2f(r)\right)\nonumber.
\eea

Let's now assume the following ansatz for $\eta_1(r,t)$
\be\label{ansatz}
\eta_1(r,t)=g(r) e^{-i \beta t}~,
\ee
and take the amplitude of the perturbation to be of the form
\beq
A_1(t) = \hat A_1(t) \cos\left( \omega_{1} t\right)\,,
\eeq
where $\hat A_1(t)$ captures the slow decay of the perturbation due to the radiation. Using the first order solution with 
$\hat A_1(0) = A_0$ in Eq. (\ref{eigen-rad-re}) we conclude that the frequency of
the radiation should be constrained to be $\beta= 2 \omega_{1}$. 
Substituting (\ref{ansatz}) in (\ref{eigen-rad-re}), and using $\beta= 2 \omega_{1}$, we get
\bea\label{eigen-rad-re1}
&&-\pa_r^2 g(r)-\frac{\pa_r g(r)}{r}+\frac{g(r)}{r^2}+\frac{1}{2}\left(3 f^2(r)-1\right)g(r)-\beta^2 g(r)=\label{re-rad1}\\
&&\frac{3}{4} A_0^2\left(\alpha s^{(1)}-\left(s^{(1)}(r)\right)^2f(r)\right)\nonumber.
\eea

Using the appropriate Green's function for the homogeneous part of Eq. (\ref{eigen-rad-re1}) we show in Appendix \ref{radiation-field}
how one can solve the previous equation to arrive at an approximate expression for the radiation field of the form
\be
\label{rad-asymptotic}
\eta_1(r,t)\approx  \hat A_1^2(t)
\left[\frac{0.0256}{\sqrt{r}} \cos\left(2 \omega_1 t -r\sqrt{4\omega_1^2-1}-\zeta\right)\right],
\ee 
where $\zeta$ is an irrelevant numerical phase.

Using this asymptotic form of the radiation we can now compute the energy flux 
in the radial direction, namely the $T_{0r}$ component of the energy momentum tensor. For the radiation we have
\be\label{energy-flux}
T_{0r}=2\dot{\eta_1}\pa_r\eta_1~.
\ee
From (\ref{rad-asymptotic}) we see that the amplitude of the radiation field is proportional to $\hat{A}_{1}^{2}$, so the 
power radiated at infinity computed from (\ref{energy-flux}) will be proportional to $\hat{A}_{1}^{4}$. In Appendix \ref{radiation-field} 
we show that the expression for the power is
\be
\dot{E}= -2\pi b_1 \hat A_1^4(t) ,
\ee
where the correct coefficient for the first mode is $b_{1}\approx0.0019$.
On the other hand, in order to determine the decay of the amplitude of the first internal mode we have to compute 
the energy associated to this mode. For a configuration of the form
\be
\phi({\bf x},t)=f(r)e^{i \theta}+s^{(1)}(r) A_1(t)e^{i \theta} ~, \label{full-exp1}
\ee
the energy can be computed directly from the $00$ component of the energy momentum tensor and it is given by

\bea\label{global-ener1}
E/\eta^2 &=& E_0 / \eta^2+ 2\pi  \int_0^\infty r dr\left(-f(r) s^{(1)}+\frac{2 f(r)s^{(1)}}{r^2}+f^3(r)s^{(1)}+2f'(r)(s^{(1)})'\right)A_1(t) + \nonumber\\
&& 2\pi  \int_0^\infty r dr\left(-\frac{1}{2}(s^{(1)})^2+\frac{(s^{(1)})^2}{r^2}+\frac{3}{2} f^2(r)(s^{(1)})^2+(s^{(1)'})^2\right)A^2_1(t)+... \nonumber\\
&\approx & E_0/\eta^2+ 2\pi \omega_1^2 \hat A_1^2(t),
\eea
where we have integrated by parts twice and used Eqs. (\ref{global-field-eq}) and (\ref{eigen-re}).

We can now equate the rate of change of the energy in the excited vortex to the power leaking towards
infinity to get
\be\label{decay-A}
\omega_1^2 \frac{d  \hat A_1^2(t)}{dt}=-b_1 \hat A_1^4(t).
\ee
Integrating this equation,
we arrive to the final result for the amplitude of the perturbation, namely,
\be\label{A1}
\hat A_1^{-2}(t)= \hat A_1^{-2}(0) + \Omega_1  t\,\qquad\quad \Omega_1=\frac{b_1}{\omega_1^2} = 0.00218\,.
\ee

\subsection{The second mode}

Let us now consider a vortex excited by the presence of the second mode, namely a field configuration described by
\be
\phi({\bf x},t)=f(r)e^{i \theta}+s^{(2)}(r) A_2(t)e^{i \theta}.
\ee
Following the exact same procedure as before we obtain a time dependent
amplitude of the form
\beq
A_2(t) = \hat A_2(t) \cos\left( \omega_{2} t\right)\,,
\eeq
where
\be
\label{A2}
\hat A_2(t)^{-2}= \hat A_2^{-2}(0) + \Omega_2  t\,\qquad\quad\Omega_2=\frac{b_2}{\omega_2^2} = 2.77 \times 10^{-7}\,.
\ee

Note that the decay time of this second mode is much longer. This is basically due to the  smaller
overlap of the source in this second mode with the profiles for the scattering modes, hence decreasing the
coupling of the source to the radiation field. A similar argument 
may be used for the higher bound modes which are in fact spread over a much larger
distance. This seems to indicate that their decay rate would also be more and more
suppressed as one approaches the continuum threshold.

The previous calculation assumes that there is no coupling between these modes and they evolve
independently. However, in a realistic situation there is always a small coupling between them. 
These couplings between modes of different frequencies will induce a variation
on the amplitude of the modes that will be superimposed on the analytic calculations presented earlier.
We will show an example of this effect in the next section.

\section{Numerical evaluation of the decay rate}

\subsection{Preliminaries}

The analytical calculations performed in the previous section rely on a set of approximations. In order to
confirm these results we have conducted a series of numerical simulations. In this section we report on
our most relevant findings. We relegate to Appendix \ref{appendix-numerics} the details of the numerical scheme that we use to simulate our
model\footnote{Many of the numerical techniques we use in this paper have also been applied in our earlier paper on 
kink excitations \cite{Blanco-Pillado:2020smt}.}. Depending on the symmetry of the situation that we want to simulate we use a $1+1$ dimensional
lattice with cylindrical symmetry or a generic $2+1$ dimensional lattice. 

Another important ingredient in our simulation scheme is the use of absorbing boundary conditions \cite{ABC}. This technique
allows us to run simulations for a long period of time without having to worry about 
the radiation bouncing off the simulation walls and affecting again the vortex. We used this type of boundary conditions 
in all our simulations except in the last section where we use periodic boundary conditions. However, the efficiency of these absorbing boundary conditions varies with the dimensionality of the 
lattice. This is due to the fact that in $2+1$ dimensions many waves will be approaching the boundary with different 
angles. This decreases the efficiency of the absorbing boundary conditions that are maximized for normal incidence. Details of our 
approach can be found in Appendix~\ref{appendix-numerics}.

In the following we will compare the evolution of the amplitude of different modes with their theoretical expectation.
In order to extract this information from the simulations we need to be able to read off this amplitude directly
from the lattice. We do that by first finding at each moment the position of the vortex in our lattice. In $2+1$ dimensions this is done 
by identifying the elemental lattice plaquette that has a non-zero winding. This is of course not needed
in  $1+1$ simulations since the vortex is always fixed at the origin by construction. After that we subtract the background 
vortex solution centered at that point from the actual field configuration. This gives us all the perturbations present
at that moment around that vortex. We can then project this perturbation field over each of the particular
excitation modes we are interested in. Using the orthogonality condition of the different modes we can 
obtain by this projection the value of the amplitude we seek. In most of our simulations this procedure yields
good results, however this needs to be refined in case the vortex is moving with some velocity.  We will
come back to this issue at the end of next section where some of the vortices are moving 
with enough velocity to make this effect relevant.

\subsection{Lifetime of the excitations}
\label{exci}

Our first simulations are aimed to check the result of the radiative decay of the 2 different internal
excitations we have numerically found in the previous section. Our initial conditions for these simulations are given by
an individual excitation of the form
\begin{equation}
\phi\left(r,\theta,t=0\right)=\left[f\left(r\right)+A_{j}\left(t=0\right)s^{(j)}\left(r\right)\right]e^{i\theta}.
\label{eq:vortex plus shape}
\end{equation}
where $j=1,2$ denote the two states we will be studying. We make use of the symmetry of the problem and run the simulations on a $1+1$ dimensional
 lattice. This allows us to run with high precision and an extremely long runtime.

We first set up the excitation separately, i.e., we either take $A_1=0.2$ and $A_2=0$, or  
$A_2=0.2$ and $A_1=0$. In both cases the extra energy stored in the internal modes for these
values of the amplitudes is about $5\%$ of the energy of the vortex core.

\begin{figure}[h!]
\includegraphics[width=8cm]{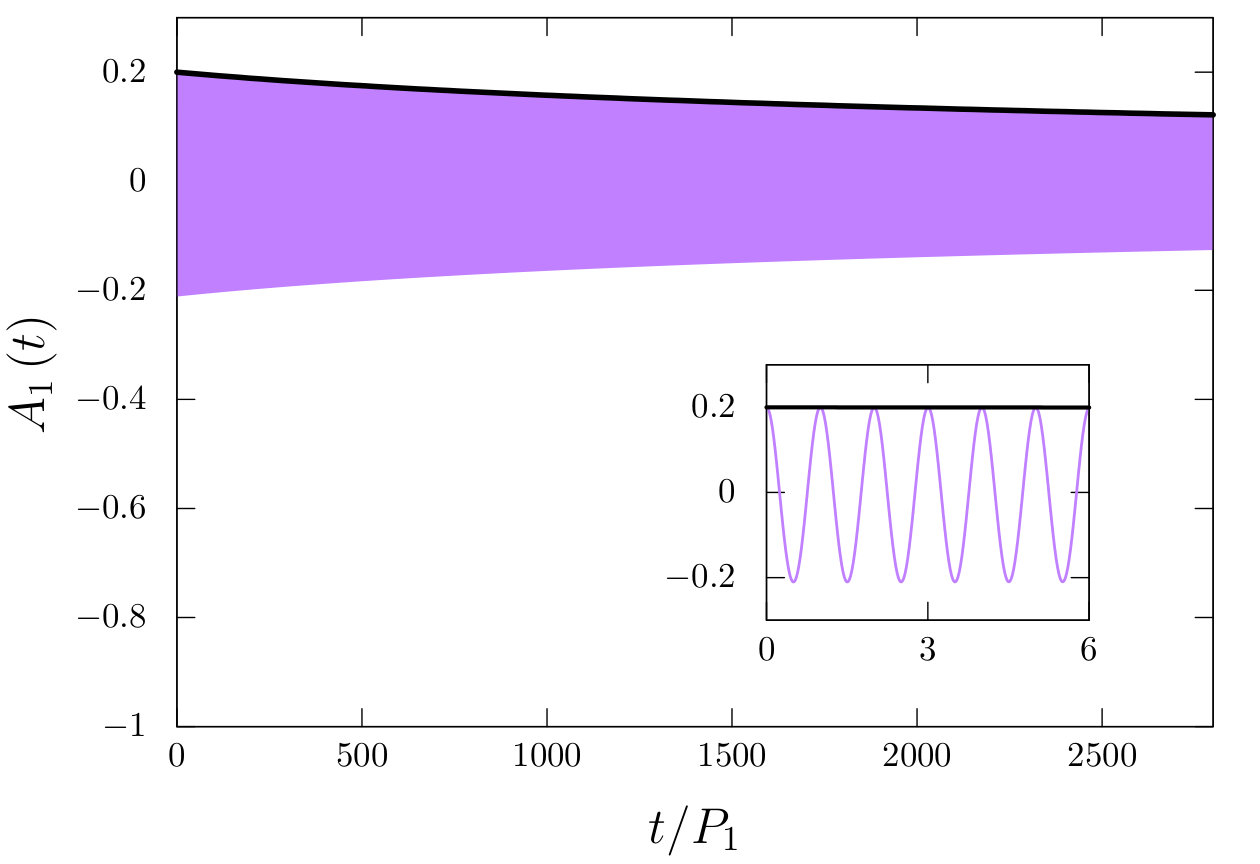}
\includegraphics[width=8cm]{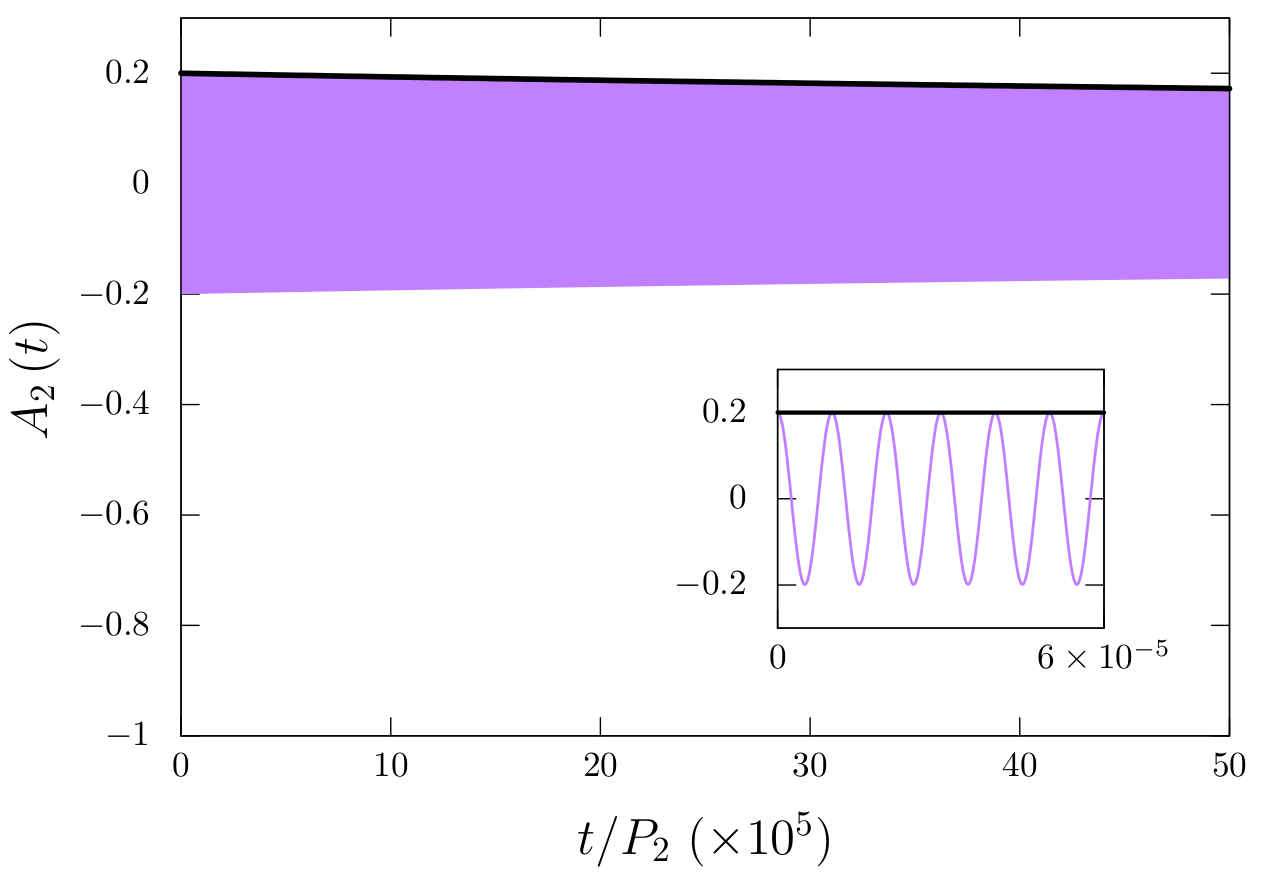}
\caption{Amplitude of the modes (purple curve) as a function 
of time (displayed in units of the corresponding period) and comparison with the 
analytical estimate (solid black line). The inset shows the first few 
oscillations for each mode. On the left we show the first mode, whose  period is $P_{1}=2\pi/\omega_{1}\approx6.97$. 
On the right we show  analogously the second mode, with period $P_{2}=2\pi/\omega_{2}\approx6.29$.}
\label{fig:amplitude}
\end{figure}

We show in Fig. \ref{fig:amplitude} the comparison between the time dependent amplitude $A_j(t)$ extracted 
directly from the simulation with the analytic predictions given by $\hat A_j(t)$. These figures demonstrate 
that the analytic calculations do indeed capture the slow non-linear decay of these perturbations. It is 
interesting to note that even increasing the amplitude beyond the linear regime, meaning taking initial 
amplitudes of the order one, the analytical expression given earlier is still a pretty good approximation. So, 
in principle, these bound states can have a large fraction of the energy of the vortex core and still behave in
a quasi-linear fashion. This means that the vortices can store quite a lot of energy 
in these internal modes for a much longer time than the natural time scale of the system, the
time associated with the width of the vortex. This property suggests the view put forward in \cite{Blanco-Pillado:2020smt}
of the excited bound states as {\it oscillons inside the soliton} type of excitations.\\

We have analysed the spectrum of the radiation obtained from the decay of the excited vortex. 
The Fourier analysis of the radiation field demonstrates that most of the energy is radiated away
at twice the frequency of the excited modes. Another interesting
point to make is that this radiation is almost exclusively produced in terms of massive 
radiation. This is somewhat expected since the original excitations have to do with the
radial modes for the scalar field. However, it is important to check that non-linear couplings
between these modes do not lead to an appreciable radiation of the Goldstone mode. This fact could
be used to disentangle the different sources of energy from a generic string state since, according to
our simulations, the coupling of the zero mode of the vortex to the Goldstone mode will lead to 
massless radiation whereas the internal excitations will
mostly emit massive modes.\\

\begin{figure}[h!]
\includegraphics[width=12cm]{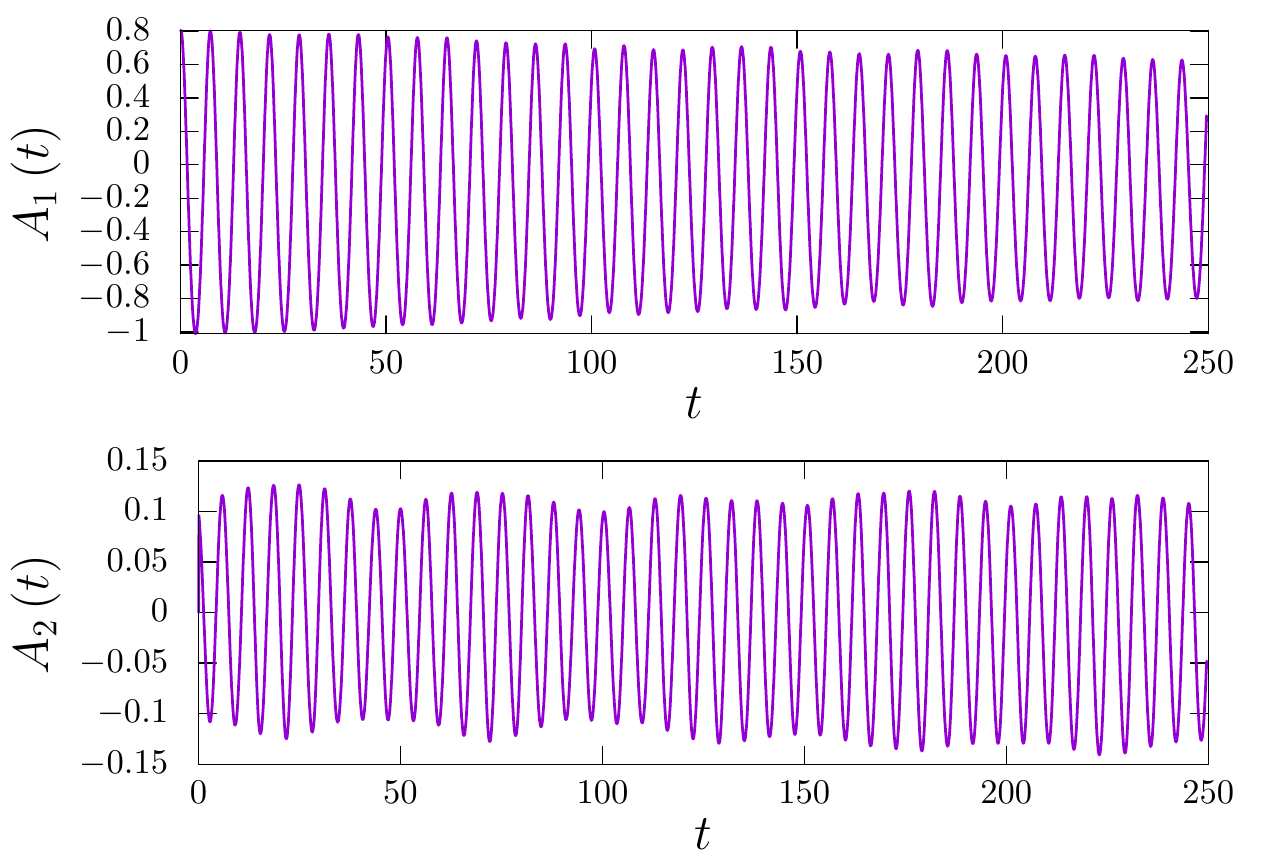}
\caption{Non-linear coupling between the bound states. The first mode is initialized with an amplitude 
of 0.8 and the second one with an amplitude of 0.1. The amplitude of both modes is modulated by
the presence of the other. The effect is more visible on the second mode.}
\label{fig:coupling bound states}
\end{figure}

We have also simulated the case where both excited states are present. In this case, the
small coupling between the modes leads to a modulated interaction between the two modes.
One can clearly see this effect in Fig. \ref{fig:coupling bound states}.

Finally, we have also checked the evolution of these modes in a less
symmetric situation by running the same initial conditions on a 
$2+1$ lattice. The results do not lead to substantial differences in the 
time dependence of the amplitude of the modes or the nature of the 
radiation emitted during the period of time that we can run in this case.

\section{Exciting the bound state modes }

In the previous sections we have studied the existence and evolution of the internal
states of the global vortex in 2 dimensions. Here we would like to explore the dynamical mechanisms that
induce such an excitation.

\subsection{Exciting the vortex with massive radiation}
\label{ilumin1}

Since the radiation of the excited states is mostly in the form of massive perturbations of the scalar field (the radial
mode in field space) it is reasonable to expect that the reverse process should also occur. Therefore, we have studied the effect of 
``illuminating" the vortex with massive radiation. In order to do that we construct an initial state in $1+1$ 
dimensions with cylindrical symmetry where we inject an incoming wave packet
towards the origin where we place the background vortex solution. After bouncing off the center, the 
reflected energy from the scattering state is absorbed in the boundary and we are left with an excited
vortex solution. Reading off the amplitude of the bound excitations on the vortex we can plot the
resultant amplitude of this scattering numerical experiment.

The specific form of the incoming wave packet is constructed as follows. As we show in Appendix \ref{appendix-perturbations},
the asymptotic form of the linearized equations of motion with cylindrical symmetry can easily be solved for this case. For the 
massive degree of freedom this solution is given by $\delta \phi(r,t) \approx J_1(k r+\omega t) e^{i \theta}$, where $J_1(x)$ 
denotes the Bessel function of the first kind and $\omega^2 = k^2  + 1$. Taking this into account we consider our initial condition to be of the form
\beq
\phi({\bf x},t) = \left(f(r) + B ~{\cal W} (r) J_1(k r+\omega t) \right)e^{i \theta}\,,
\eeq
where we have taken ${\cal W}(r)$ as a window function that gives support to the
wave on a finite region of space in our simulation, initially away from the core of the
vortex. We take this window function to be
\beq
{\cal W} (r) =\frac{1}{4}\left[1+\tanh\left(r-20\right)\right]\left[1-\tanh\left(r-50\right)\right]~.
\eeq

\begin{figure}[h!]
\includegraphics[width=10cm]{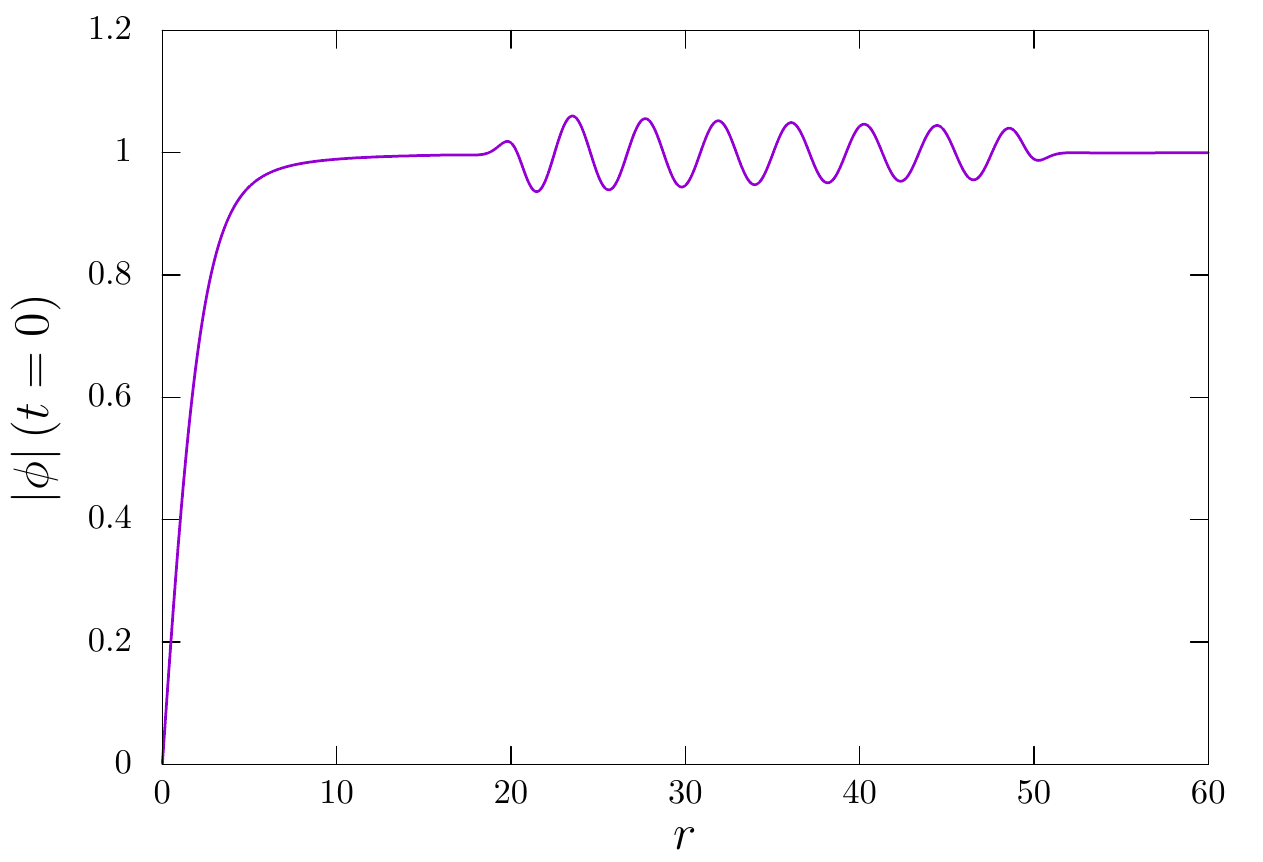}
\caption{Radial part of the field at $t=0$. The incoming wave has amplitude $B=0.46$ and angular frequency $\omega=2\omega_{1}$.}
\label{fig:massive radiation initial}
\end{figure}

An example of the initial state is shown in Fig. \ref{fig:massive radiation initial} where we have taken 
$B=0.46 $ and $\omega= 2 \omega_1$, where $\omega_1$ is the frequency of the
first bound state.

\begin{figure}[h!]
\includegraphics[width=10cm]{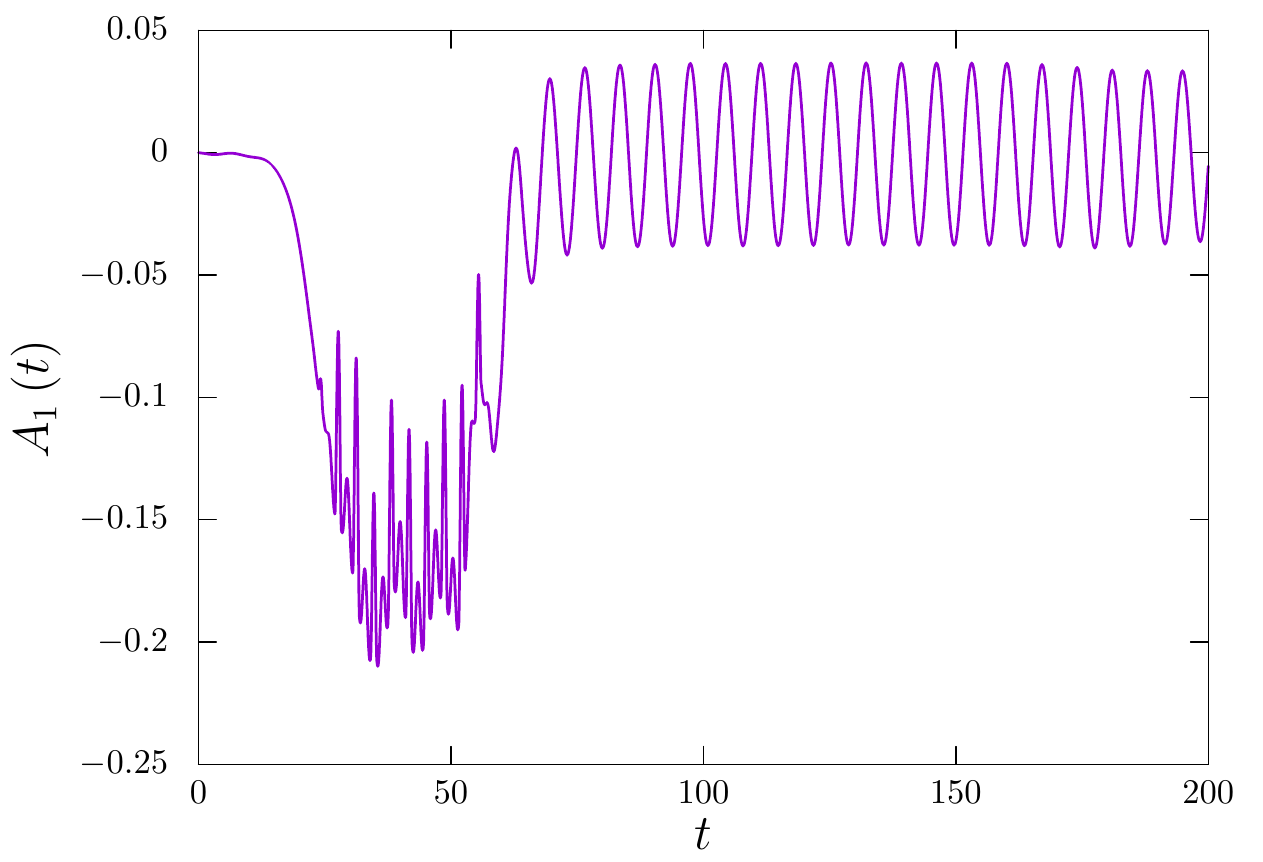}
\caption{Amplitude of the first bound state as a function of time. After an interval of overlapping with the vortex, situated at $r=0$, the 
wave recedes to infinity (being absorbed at $r=60$) leaving the bound mode excited.}
\label{fig:massive radiation ampl}
\end{figure}

For this particular initial configuration, the amplitude of the first bound state as a function of time is 
shown in Fig. \ref{fig:massive radiation ampl}. As we anticipated, the bound state is excited 
in this process. However, the level of excitation depends on a complicated way on the frequency as well
as the amplitude of the incoming wave, and does not seem to follow a simple pattern. This fact seems to
resemble the results found in $1+1$ dimensions for the case with the kink soliton in \cite{Romanczukiewicz:2005rm} where 
a fractal-like behaviour was encountered. Our purpose here is just to identify the different physical
processes that can excite the bound state for the vortex, so we will not study this in more detail.

We have also run similar experiments in $2+1$ dimensions to make sure that the effect does not
depend on the restricted symmetry of the radiation. In this case, we use a plane symmetric wave colliding
with the vortex. The result of this scattering process is complicated by the fact that the vortex
is displaced by the incoming wave. In other words, it is not only the bound state that is excited but also
the translational zero mode. 

These numerical experiments have also uncovered an interesting effect due to the non-linear
interaction of the radiation and the vortex. As the massive wave passes by the vortex, this reacts by
moving in the direction of the incoming wave. This may sound counterintuitive but in fact is
a process that has already been observed to happen in the case of kinks in $1+1$ dimensions
and has been dubbed in the literature as negative radiation pressure \cite{Forgacs:2008az, Romanczukiewicz:2003tn, Romanczukiewicz:2005rm, Forgacs:2013oda, Romanczukiewicz:2008hi}.

\subsection{Exciting the vortex with massless radiation}
\label{ilumin2}

We are also interested to see whether one can excite the first massive mode localized on the vortex by 
illuminating it with a wave of the massless field, the Goldstone mode. At the linear level this does not seem 
to be possible; however, at the non-linear regime, it could happen. 

\begin{figure}[h!]
\includegraphics[width=10cm]{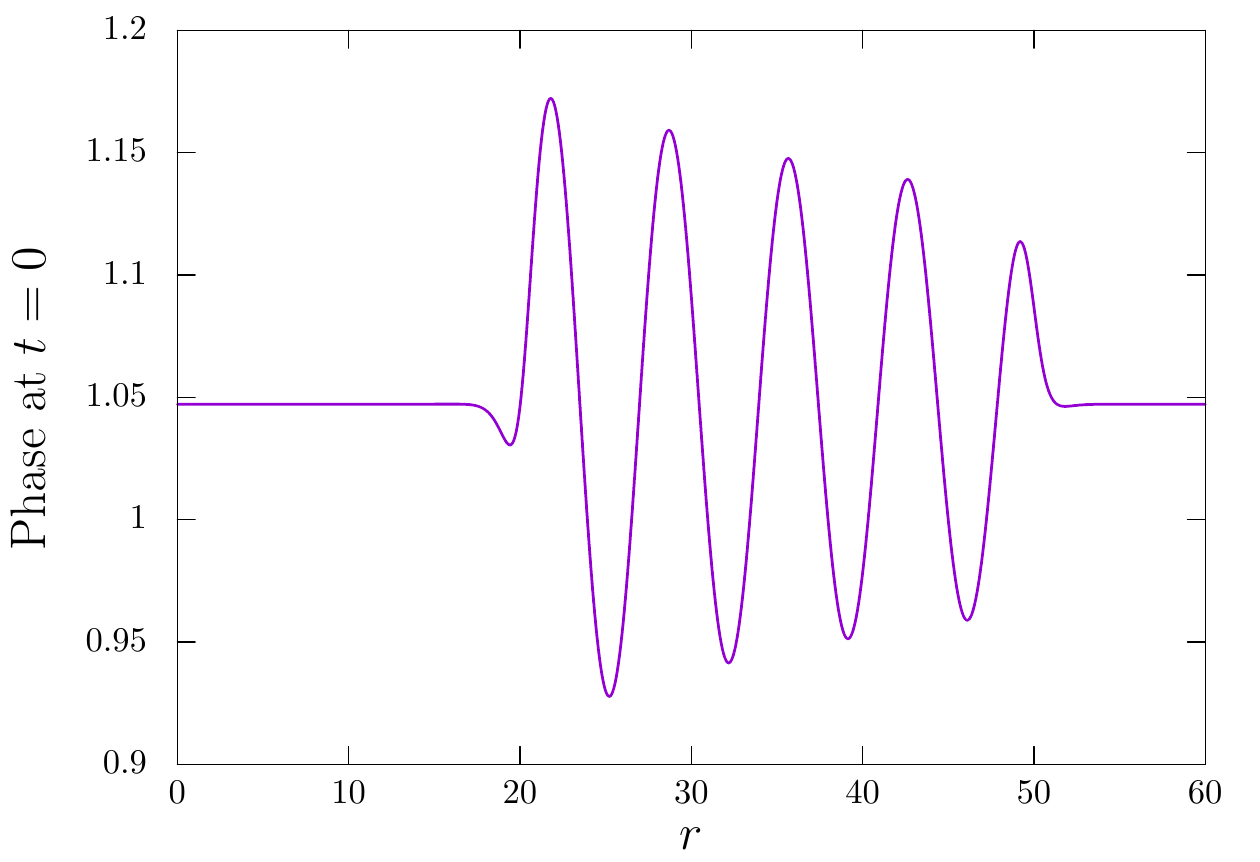}
\caption{Phase profile for the incoming wave given by Eq. (\ref{massless-wave-profile}) with amplitude 
$B=0.71$ and angular frequency $\beta\approx\omega_{1}$.}
\label{fig:massless radiation initial}
\end{figure}

\begin{figure}[h!]
\includegraphics[width=10cm]{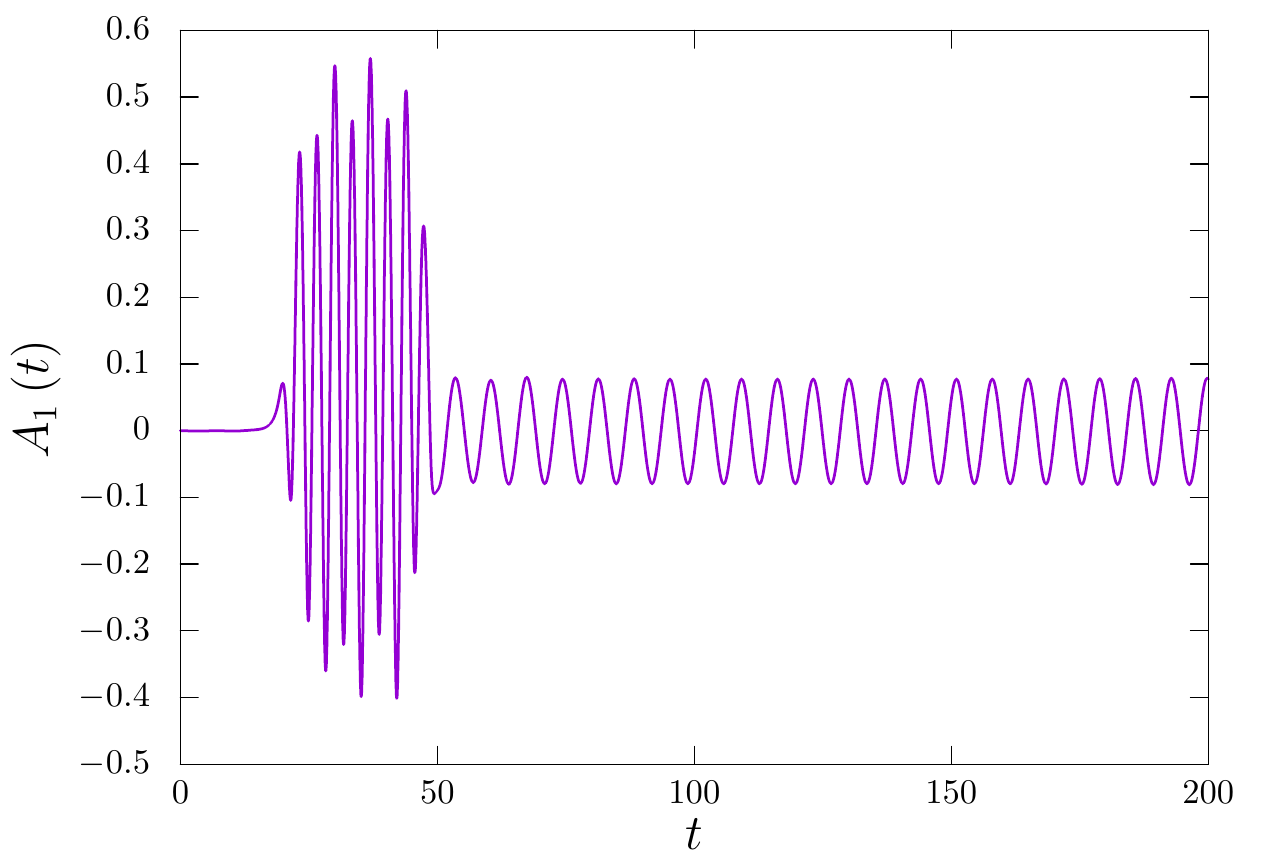}
\caption{Amplitude of the first bound state as a function of time. The massless wave interacts 
with the vortex and leaves the bound mode excited.}
\label{fig:massless radiation ampl}
\end{figure}

The experiments we have performed in this case are pretty much the same as in the previous
section. We first radiate the vortex with a wave packet constructed from scattering states of the 
massless field in a cylindrical fashion. Following the description given in the previous section and the
discussion in Appendix \ref{appendix-perturbations}, one can show that such perturbation can be described by a field configuration
of the form

\beq
\label{massless-wave-profile}
\phi({\bf x},t) = \left(f(r) + i B ~{\cal W} (r) J_1(k r+\omega t) \right)e^{i \theta}\,,
\eeq
where once again $J_1(x)$ denotes the Bessel function of the first kind. Such initial state is shown in Fig. \ref{fig:massless radiation initial} for
a particular set of parameters. The results of the interaction of this wave with the 
vortex is shown in Fig. \ref{fig:massless radiation ampl} where we display the amplitude of the first bound state.

 This is an interesting result since it clearly shows that an incoming massless radiation can excite these internal modes. This effect could
have important consequences for numerical simulations of string networks where the motion of the
strings creates a background of massless radiation that could transfer part of its energy to these
internal modes by this long range interaction of vortices.

Similarly to the case with the massive wave, the amplitude of the bound state at the end
of the scattering process depends on the frequency as well as the amplitude of the wave
in a complicated way. We have not tried to systematically study this dependence in any 
detail.

We have also investigated this effect in less symmetrical situations where we radiated
the vortex with a plane wave of the massless mode. The result of these investigations
also show that the vortex becomes excited by the interaction with these waves so we conclude that
this is a generic effect.

Interestingly, the interaction with these waves also displaces the vortex not only in the direction of
motion of the wave but in the transverse direction. This transverse motion can
be understood from the point of view of the dual picture as the interaction of a
charged particle (the vortex) with an electromagnetic wave (the massless Goldstone 
wave).

Finally, we notice that, in the numerical examples we have performed for this case, we do not
see the negative pressure effect. This seems to distinguish the effect of
massless and massive radiation on the vortex in this regard.

We have not performed the analogous tests for the second bound state. Its large
size suggests that it would only be excited by waves of similar wavelength. This makes
these simulations computationally costly. We will see later on in our cosmological simulations that this mode
is indeed also excited, so we expect that the same kind of processes as the
ones shown here take place for that mode in a realistic setting. An analogous comment should apply for bound 
modes above $s^{(2)}(r)$. In an ideal situation, they would be excited by even larger wavelengths but in practice, they will 
be suppressed by the presence of a distance cutoff; either the size of the simulation box or by the distance between
the vortices.  \\\\

Before concluding this section we would like to comment on an interesting effect that
we have also observed in numerical simulations. Apart from looking at the possible
excitation of the bound states from waves scattering off the vortex we have also
explored the effect of these waves on a vortex with an initial state with a non-zero amplitude
for a bound state, in other words, on a vortex in an already excited state. In the course of these investigations we have noticed that
a vortex with a large amplitude mode that is irradiated with waves does
not seem to absorb more energy in the bound state. In fact, in some cases, the system 
seems to decay faster than it would have done in the vacuum. This sort of {\it stimulated emission}
process could have important consequences to set the typical level of
excitation in vortices in realistic situations. We will comment on this possibility
in the next sections.

\subsection{Vortex-antivortex interaction}
\label{vavi}

As we explained earlier, one can describe the low energy dynamics of a $2+1$ dimensional
global vortex with an effective action similar to the well known Kalb-Ramond action \cite{Hecht:1990mv,BlancoPillado:2009di,Fleury:2016xrz}.
In this case, the Goldstone mode is replaced by its dual formulation which in $2+1$
dimensions is a Maxwell field. Vortices are therefore replaced in this effective theory
by point-like particles of fixed mass and charged with respect to the gauge field.

It is then easy to understand in this language why vortices would attract or repel each other
depending on their relative winding. Radial modes are by definition not described in this 
effective theory. However, the acceleration of vortices moving under the influence of another 
nearby vortex (or anti-vortex) can lead to the excitation of the internal modes. 

We have indeed observed such behaviour in pairs of vortices that we placed initially at a
distance larger than their core sizes and that were gradually accelerated away from each other.
A quantitative measure of the amplitude of the bound state is complicated by the fact that
in this experiment the vortices acquire a large velocity. One could in principle numerically
transform our lattice data into the rest frame of the vortex of interest. This would give us the field distribution that an 
observer moving at the instantaneous velocity of the vortex would see and read the amplitude
from there. This is however quite a challenging procedure numerically, specially when we have 
large variations of the velocity in short periods of time. In this paper we will only consider 
vortices with small velocities where we have checked that this issue does not affect the 
numerical estimate of the bound state's amplitude.

\section{Bound state excitation at formation}
\label{at formation}

Previous numerical experiments show that the internal states of a vortex can be excited by absorbing part of the
radiation impinging on them or by interacting with other vortices. All these effects would be
present in realistic situations possibly creating a sizeable amplitude for these modes. In the following 
we will describe the numerical simulations we have performed to identify the characteristic
amplitude of these bound states when they are formed. In particular, we will study a 
couple of situations that could be relevant for the formation of vortices in cosmology or condensed matter systems.

\subsection{Vortex formation during a phase transition}

\begin{figure}[h!]
\includegraphics[width=16cm]{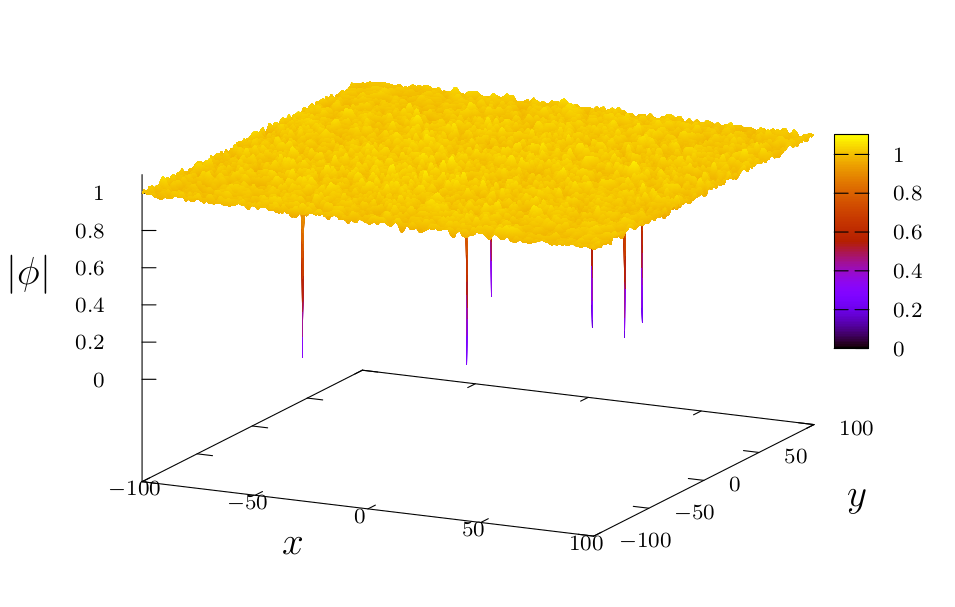}
\caption{Vortices formed in a phase transition. The color palette indicates the modulus of the field. We can clearly
identify the presence of a few vortices embedded on a background of small amplitude fluctuations
of the field coming from the phase transition.} 
\label{fig:vortices after phase transition}
\end{figure}

We are interested in studying the formation of vortices in a $2+1$ dimensional phase transition.
In order to simulate this process we will assume that the effective potential describing the dynamics
of the complex scalar field changes abruptly at some particular moment and we simulate the subsequent
formation of vortices. Starting with an initially parabolic potential we consider a thermal initial state for the 
massive degrees of freedom. After the transition we consider the potential described in our
original model in Eq. (\ref{themodel}). The initial symmetric state with $\phi({\bf x}) \approx 0$ rolls down 
the potential influenced by the thermal perturbations leading to the formation of vortices and antivortices 
{(see Fig. \ref{fig:vortices after phase transition})} \footnote{For the details of the numerical implementation
of this phase transition please look at the description in Appendix \ref{lattice-thermal-state}.}.

In our attempt to emulate a realistic setting of vortex formation we introduce a source of dissipation in
our simulations. Following what was recently done in a similar study in a $1+1$ field theory \cite{Blanco-Pillado:2020smt}, we 
will simulate the formation of vortices in a $2+1$ dimensional de Sitter space. There are several
advantages in using this procedure. First, an expanding spacetime introduces a natural friction that
allows the excess of background energy to be depleted as time passes. Furthermore, the existence
of a horizon distance precludes all the vortices to travel large distances and annihilating with
each other. Finally, and very importantly for us, setting a Hubble rate small enough, this expanding
background will not affect the oscillating bound state modes. All these are desirable properties
for our purposes and allow us to finish each of the runs with a set of excited vortices and antivortices.
We end the simulation by smoothly transitioning to a Minkowski background where there
is no expansion so we can use the previously discussed projections and read off the amplitude of the
bound states.
The price to pay for simulating an expanding background in a lattice is well known. The simulation
takes place on a comoving lattice so a fixed size physical object, as for example a vortex, will
decrease its size on the comoving lattice. We should therefore make sure that our expansion runs for
long enough to do its job of smoothing the background but not so much that our vortex core is
not well resolved. 
 
We have run $10$ different initial configurations with the thermal state and obtained the amplitude
of the internal excitations for both modes for each of the vortices present in the lattice at the end of the
simulation. Each of the realizations was done using a $2+1$ dimensional lattice with $N=2000$ points
on each direction with $\Delta x=0.1$ and $\Delta t=0.04$. We give in Appendix  \ref{appendix-numerics} and \ref{lattice-thermal-state}
the details of the numerical implementation of the simulation.

The typical number of vortices per realization is of the order $\sim 7$ and their final velocities were very small. Therefore,
we are confident that we have been able to correctly extract the level of excitation in each mode.
Averaging over all the vortices in our ensemble, we arrive to the following results for the amplitude
of the first two modes
\beq
\langle\hat{A}_{1}\rangle=0.164\pm0.016\,,
\label{amplitude 1 phase transition}
\eeq
\beq
\langle\hat{A}_{2}\rangle=0.181\pm0.060~.
\label{amplitude 2 phase transition}
\eeq

In order to estimate how excited these vortices are, we compare the extra energy
stored in these modes with the energy the vortices would have in their absence.
We do this in two different ways. We first compare the extra energy stored
in the bound states with the energy in the core. Following this prescription we
find that a vortex with the amplitudes given in Eqs. (\ref{amplitude 1 phase transition}- \ref{amplitude 2 phase transition})
has of the order of $3\%$ and $4\%$ higher energy than the core mass of the vortex in the minimum energy configuration.

We can also consider the total amount of energy in the vortex where we also include the component
associated with the gradient of the phase. This is complicated by the fact that this energy component is divergent
and therefore our result will depend on the cutoff we introduce (See the discussion in Section (\ref{section-string-background})). In our case we will consider the average
distance between vortices as the natural cutoff for the energy calculation\footnote{This is different than in the
case of domain walls (kink solutions) or local strings where the energy density of the
solitons in those cases is much more localized in their core.}. This computation in our
case shows that the vortices seem to have a $0.4\%$ and $0.5\%$  of extra amount of energy
due to these localized excitations.\footnote{As we mentioned earlier the size of the bound states 
beyond the second mode makes the analysis of their excitation during the phase transition numerically 
unfeasible. However, we do not expect any dramatic effect for those modes in any realistic 
situation.}

This seems to indicate that even though vortices get created in an excited state this extra energy
may not be too relevant. It is also interesting to note that this amount of energy
is quite lower than what was found for the case of the kink in \cite{Blanco-Pillado:2020smt}.  In the following we will also explore what happens
to this level of excitation in the course of the subsequent evolution of the vortices in a cosmological
background.

\subsection{A vortex on a thermal bath.}

\begin{figure}[htb!]
\includegraphics[width=12cm]{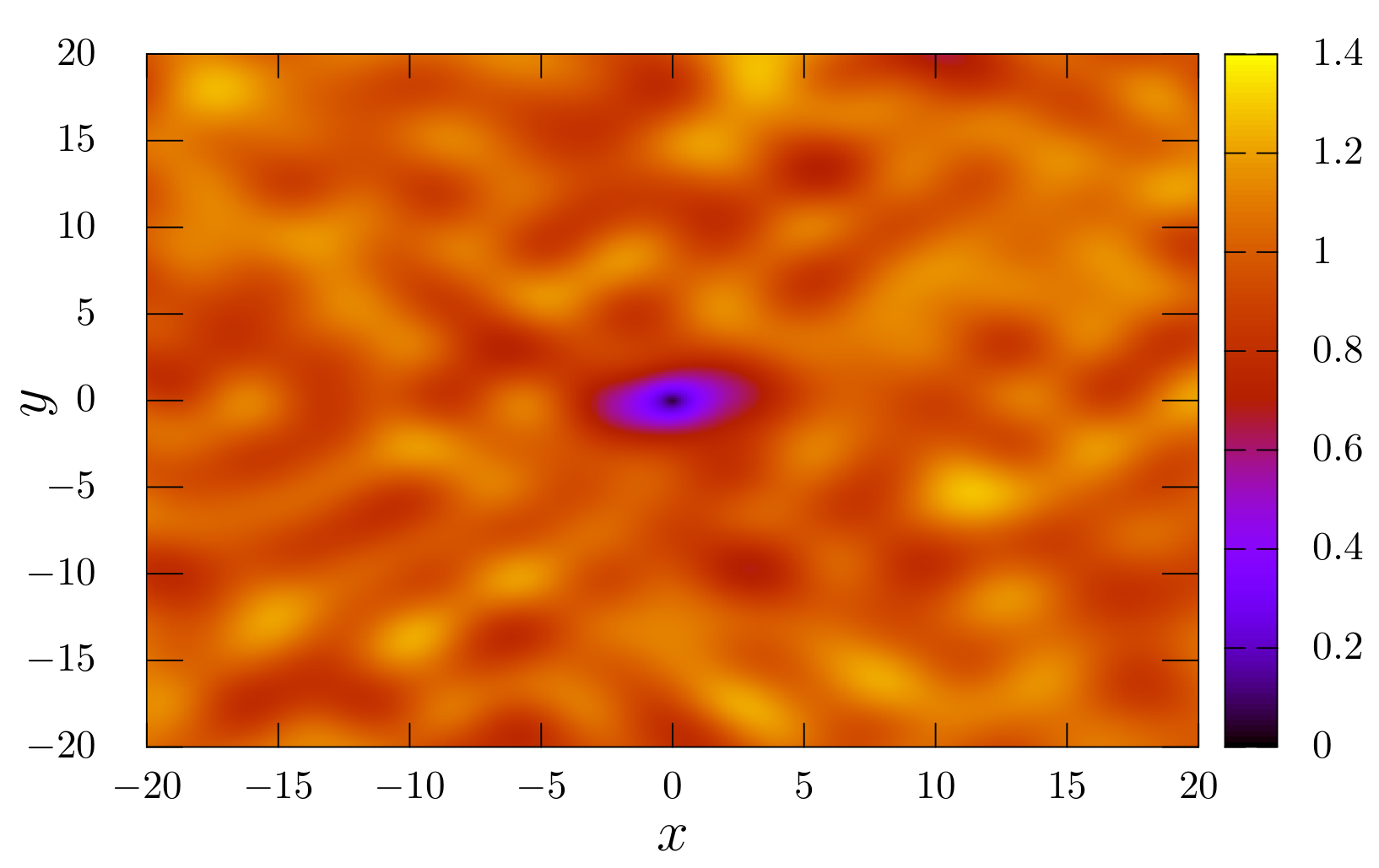}
\caption{Vortex in contact with a thermal bath at temperature $\Theta = T/\eta^2=0.1$. The color 
palette indicates the modulus of the field.}
\label{fig:thermal vortex}
\end{figure}

As we showed in a previous section, the interaction of the vortex with radiation induces the 
excitation of the bound states. In this section we would like to study the amplitude of these
modes when the vortex is placed in contact with a thermal bath (see Fig. \ref{fig:thermal vortex}). 
We give a detailed description of the numerical methods to implement this initial thermal state
in our lattice in Appendix \ref{lattice-thermal-state}. This type of configurations will be interesting  in the study of defect formation 
on reheating scenarios \cite{Tkachev:1995md} and possibly in other realistic situations in condensed 
matter physics. 

\begin{figure}[h!]
\includegraphics[width=8cm]{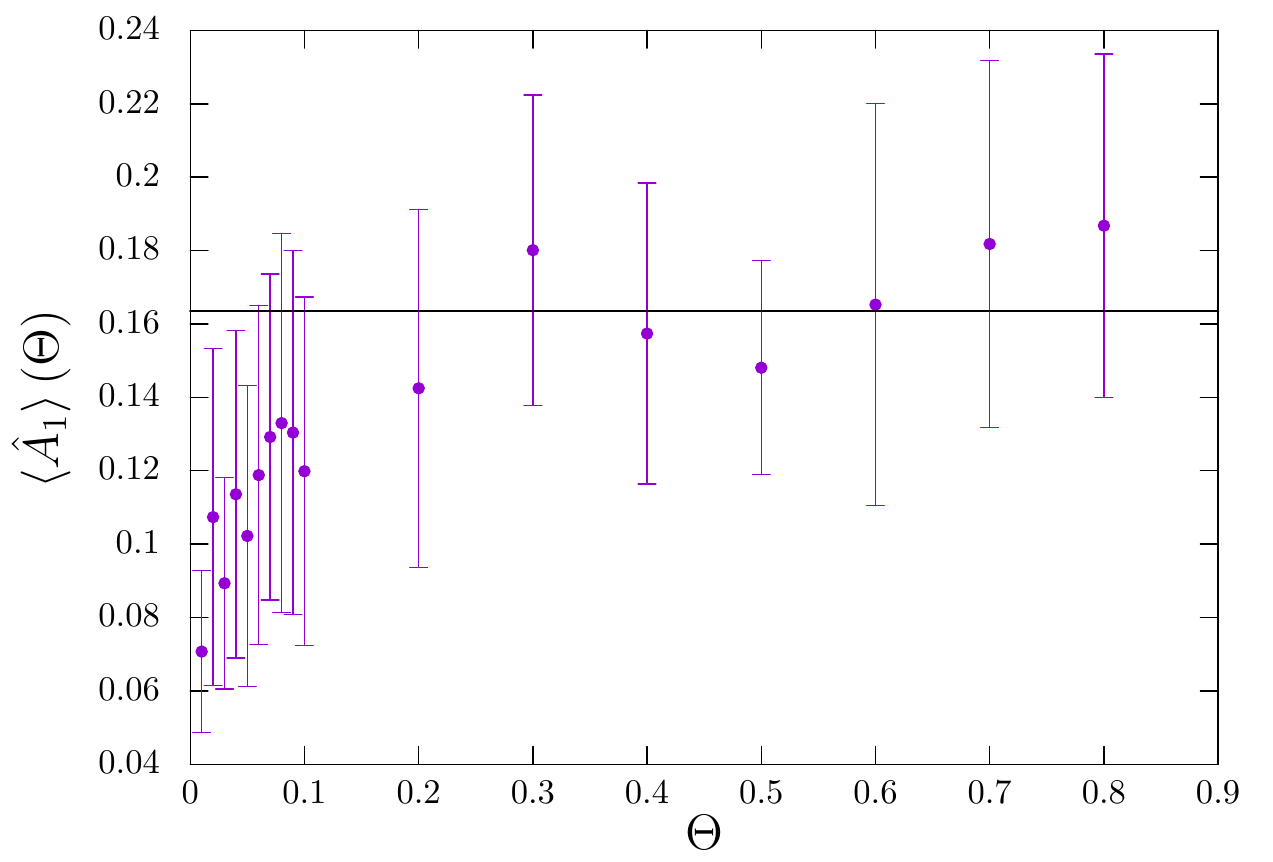}\includegraphics[width=8cm]{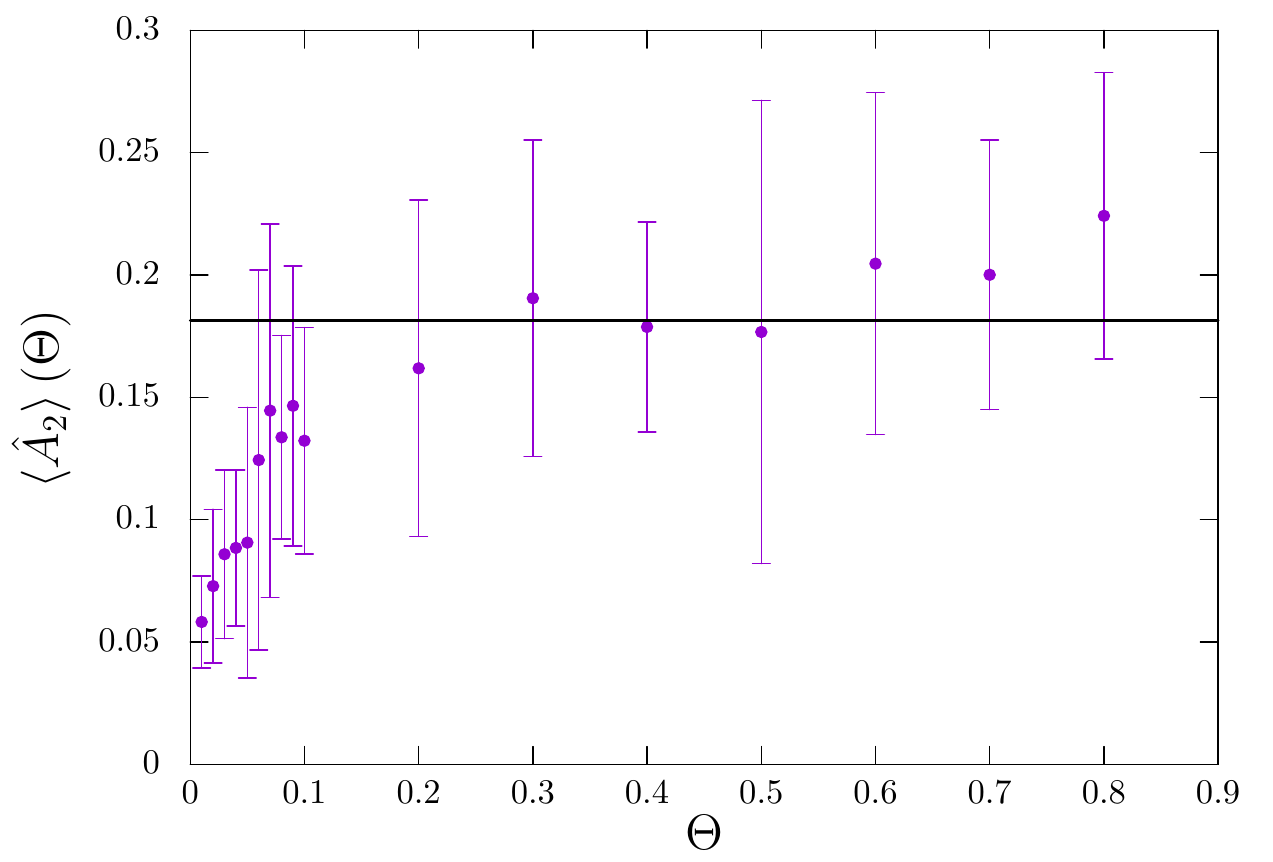}
\caption{Average amplitude of the first (left) and second (right) modes, $\langle\hat{A}_{1}\rangle$ and $\langle\hat{A}_{2}\rangle$ respectively,  as a 
function of the dimensionless temperature $\Theta = T/\eta^2$ . The black line corresponds to the average amplitude obtained in the 
phase transition, namely Eqs. (\ref{amplitude 1 phase transition}) and (\ref{amplitude 2 phase transition}) respectively.} 
\label{fig:amplitude temperature}
\end{figure}

On the other hand, this thermal bath of perturbations will also induce a random velocity to the vortex. This limits our
ability to read off the amplitude of the bound state for long periods of time since the vortex may
end up leaving the simulation box. In order to ameliorate this problem we run the simulation on
a $2+1$ dimensional de Sitter background. We choose the Hubble constant of this spacetime to
be small enough so it will not affect the amplitude of the bound states but large enough so the
induced horizon distance is within our box. This ensures that we are able to read the amplitude of the vortices after their
initial ``peculiar" thermal velocities have been redshifted away. In Fig.~\ref{fig:amplitude temperature} we 
can see  the dependence of the amplitude of the bound states with respect to (dimensionless) temperature.

We note a similar effect to the one we obtained in the numerical study of $1+1$ dimensional kinks in 
 \cite{Blanco-Pillado:2020smt}  where the amplitude of the internal states saturates at high temperature at 
 approximately the same value that is found when the soliton is formed in a phase transition.  This makes 
 sense since at very high temperatures the amplitudes of the
thermal perturbations are large enough to almost climb to the top of the potential. In this
limit, there is no much difference between this initially very hot state and the state before
the phase transition, that is why the results are consistent with one another. In fact, at
those very high energies, we have seen the formation of other vortex-antivortex pairs
from the vacuum itself as one would expect if there is enough thermal energy available.

One may wonder why there seems to be this saturation effect so the amount
of energy stored in the bound state seems to be cap at a somewhat low value compared to the
maximum possible one. One reason for this may be the existence of this process of stimulated 
emission that we mentioned in the previous section. This process will make high amplitude
states much more unstable than one would imagine studying their decay in vacuum. This
would effectively set a lower maximum amplitude for the bound states in realistic setups
where the vortices are typically hit by waves of different frequencies and amplitudes.

\section{Cosmological evolution of a $(2+1)$d ``vortex network"}
\label{cosmoevol}

As we explained earlier, the interaction of the vortex with the massive and
massless radiation seems to induce its excitation. Furthermore, their motion
due to the interaction with other vortices accelerates them, what also leads
to an excitation. It is therefore clear that the amplitude of a bound state of a vortex will not
only depend on the initial conditions but also on its subsequent dynamics. We would
like to have some idea of the evolution of these excitations as a function of time when
all these previously discussed effects are at play. 

In order to do that we will take inspiration in the type of evolution we expect to 
occur for global strings, the $3d$ counterparts of our vortices, and we will simulate 
an ensemble of vortices for a period of cosmological expansion.
In particular, we will consider a $2+1$ dimensional expanding universe
in a radiation dominated state. This type of simulation has been done 
before also as a toy model to understand the dynamics of global strings in 
\cite{Yamaguchi:1998iv}. There are however important differences with the $3+1$ dimensional
case that make a direct extrapolation between these models difficult. We will comment
more on this fact in our conclusion section.

\begin{figure}[h!]
\includegraphics[width=8cm]{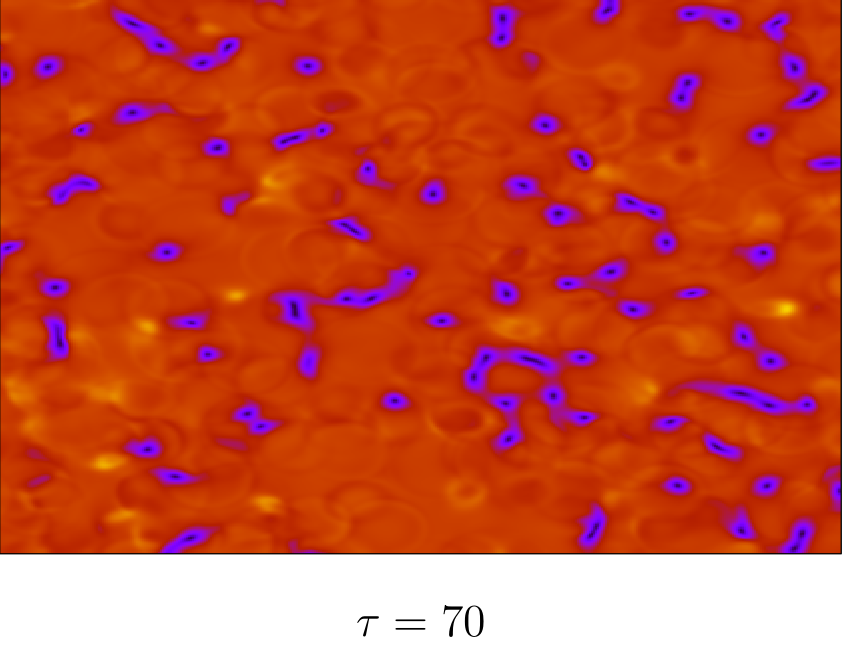}
\hspace{0.2cm}
\vspace{0.8cm}
\includegraphics[width=8cm]{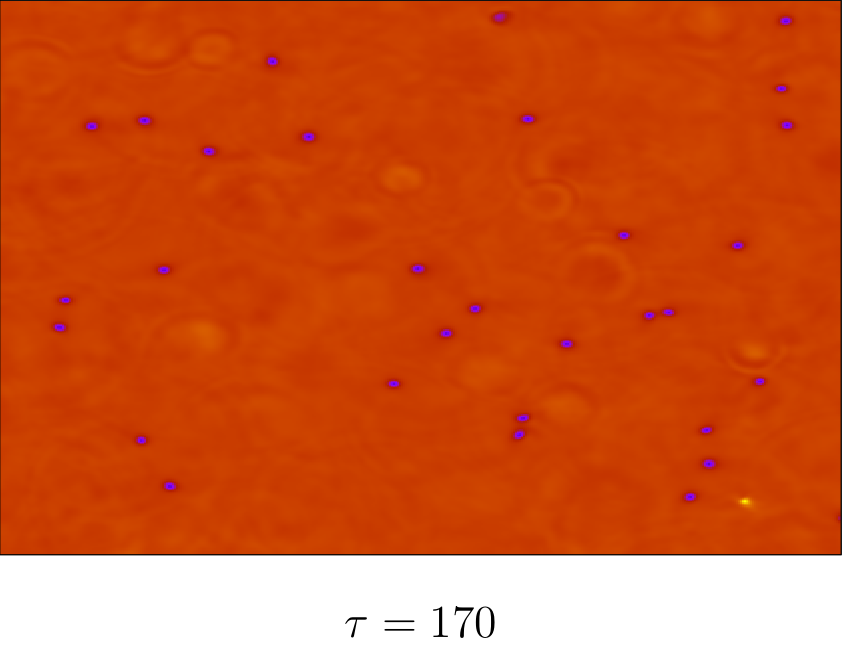}
\includegraphics[width=8cm]{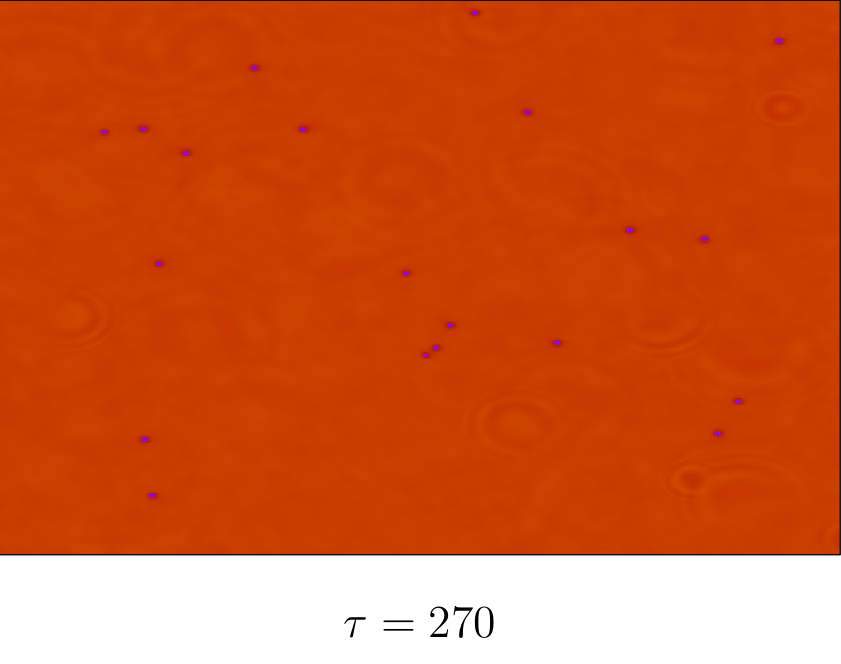}
\hspace{0.2cm}
\includegraphics[width=8cm]{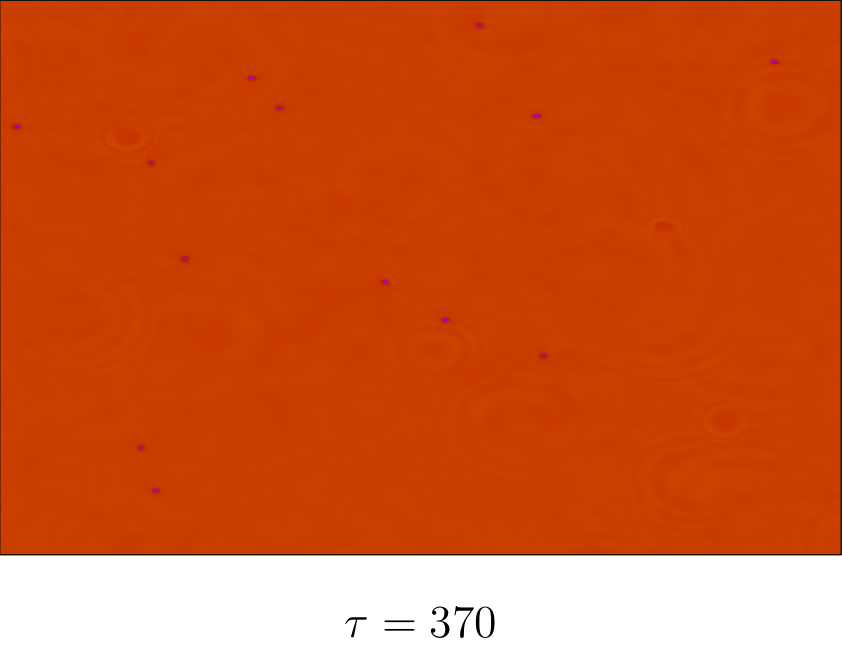}
\caption{A sequence of snapshots of the scalar field evolution during the initial part of the radiation era 
in one simulation.  The pictures show a subregion of side  $L/2$ of the simulation at different values
of the conformal time, $\tau$.} 
\label{fig:FT-snapshots}
\end{figure}

Our initial conditions are prepared following the prescription given in \cite{Hindmarsh:2019csc} 
applied now to our $2+1$ lattice (see Appendix~\ref{simcos}). This procedure has 
a couple of important regimes that are designed to end up with a convenient initial state for 
the physical evolution during the cosmological expansion. In particular, we have a period of diffusion
to get rid of much of the extra energy in the system as well as a period of core-growth where
the equations of motion are modified to quickly give the vortex the correct
size for the intended initial scale factor \footnote{For details of these regimes, see \cite{Hindmarsh:2019csc}.}. 
In our case, the end result of this procedure gives us a collection of well separated vortices that start their cosmological
evolution with a horizon distance larger than the size of the vortex. Furthermore,
the number of lattice points per vortex width is very large at this point in order to
account for the shrinking of the comoving vortex size in the course
of the simulation.\footnote{Note that in this work we are interested in getting an idea of how 
the interplay of all the mentioned processes could give rise to a network of excited vortices. 
It would be interesting to explore how the results presented here change by
using different parameters or even a different technique to set up the initial conditions of the simulation. 
This would require a dedicated study on the matter. We have purposely adapted the same procedure to find the
initial conditions for the network as is done in the $3+1$ dimensional cases
in order to facilitate a possible comparison between these two situations.}

We then evolve this configuration in a radiation dominated universe for a period
of time so that our final scale factor has grown by $a(t_f)/a(t_i) \sim 10$.
The horizon distance grows during the cosmological evolution but it is always small enough that
the periodic boundary conditions we use are not a relevant issue.

During this evolution all the processes detailed earlier do take place in our simulations. See the sequence of
snapshots at different stages of the simulation in Fig.~\ref{fig:FT-snapshots}\footnote{A movie of this simulation 
can be found in  \url{http://tp.lc.ehu.es/earlyuniverse/global-vortex-simulations/} .}. Vortices move under
the influence of other nearby ones and accelerate acquiring in some cases large
velocities. In this motion they emit radiation in the form of massless (Goldstone) modes. Some 
vortex-anti-vortex pairs annihilate sending a shell of massive radiation\footnote{We have not witnessed the
appearance of 2 dimensional oscillons as a result of these annihilations. This seems to be
in agreement with the findings in \cite{Gleiser:2007te}.}. All these waves of radiation affect other
vortices and excite their bound state modes. These processes continue until the energy
in these waves is redshifted away by the expansion of the universe.

\begin{figure}[h!]
\includegraphics[width=10cm]{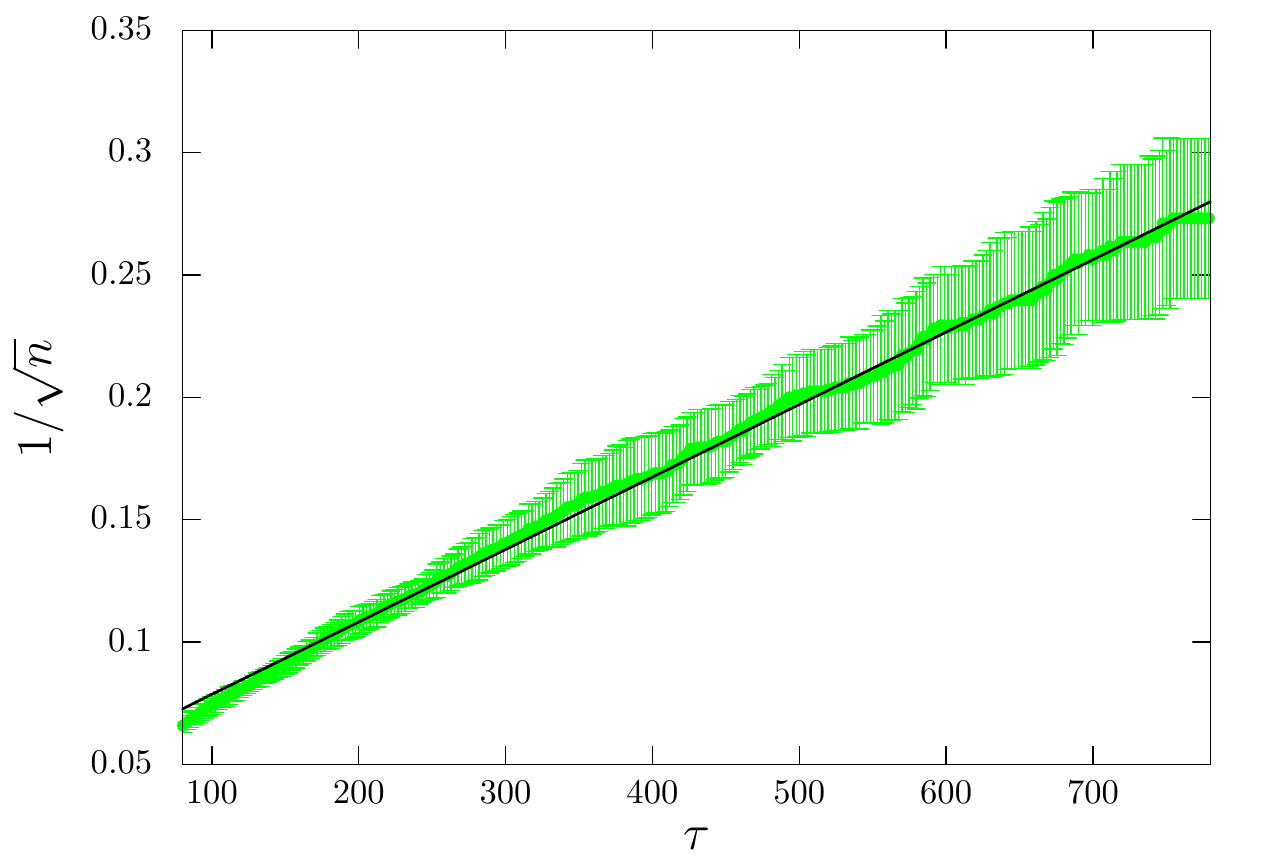}
\caption{Inverse of the square root of the average number of vortices ($n$) as a function of conformal
 time. The results are in good agreement with the linear fit $0.05+0.0003\tau$ (black line).}
\label{fig:number-of-vortices-per-Hubble}
\end{figure}

Similarly to what happens in $3+1$ dimensional networks of strings, our configurations reach
a scaling regime where the number of vortices per Hubble length is roughly constant \footnote{We do not concern ourselves
here with the possible $\log$ dependence since we are not interested in this issue. Specially taking into account that
$2+1$ dimensions is in fact special in this regard \cite{Yamaguchi:1998iv}.}. This in turn means that the number of vortices
should go as $n(\tau) \propto \tau^{-2}$, where $\tau$ is the conformal time in our cosmological simulations. 
We show in Fig. \ref{fig:number-of-vortices-per-Hubble} the behaviour of the average number of vortices 
for our simulations. As the figure shows, our simulations enter a scaling regime pretty early so we are able 
to run within this phase for a substantial amount of our run-time thus also factoring in the
possibility that internal modes evolve during this period. Finally, in order to read off the amplitude of the bound states 
we finish the cosmological evolution of the scale factor by smoothly transitioning to a flat space background.

\begin{figure}[h!]
\includegraphics[width=8cm]{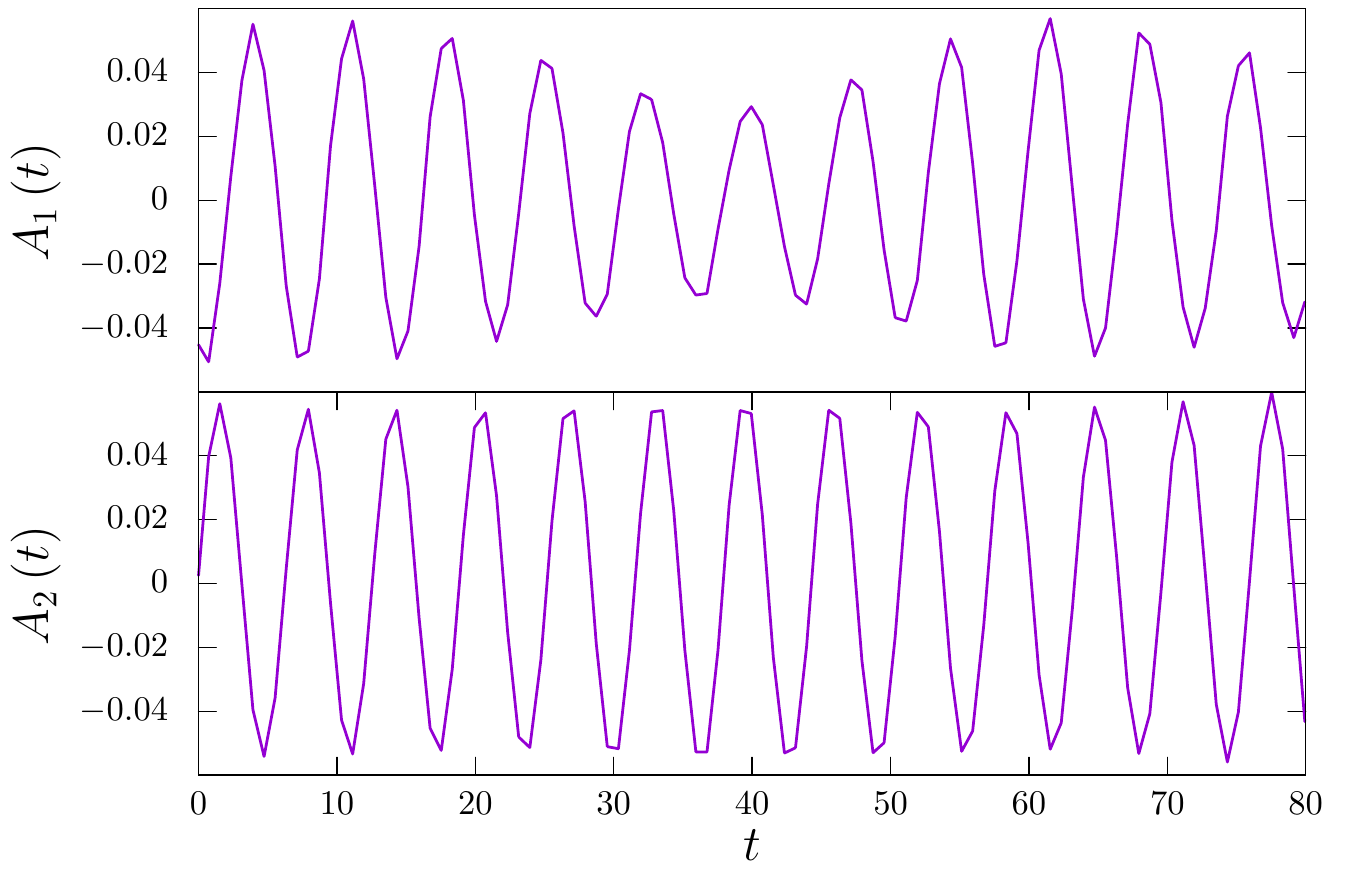}
\includegraphics[width=8cm]{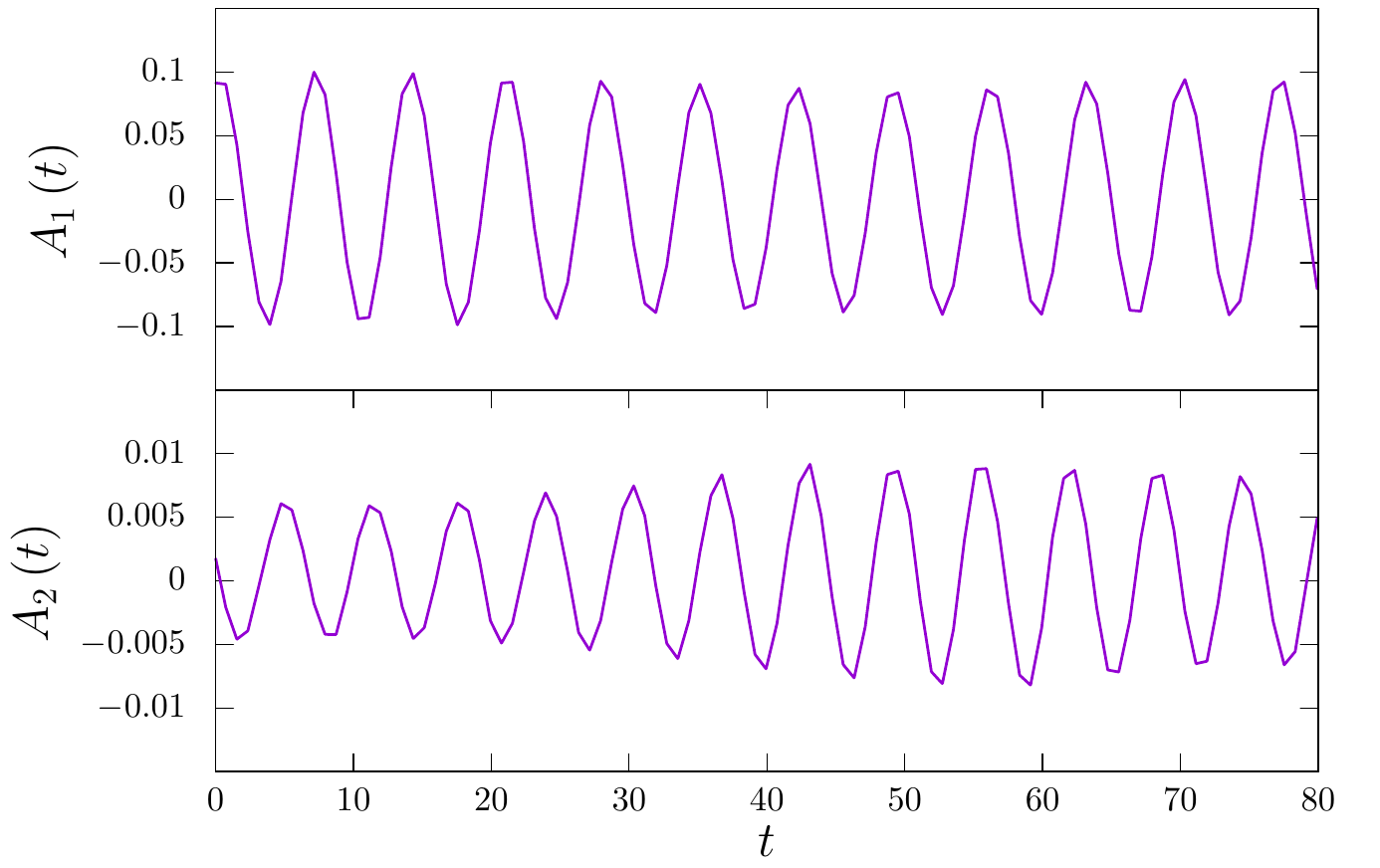}
\caption{Two examples of the amplitude of the bound state modes on a vortex at the end of the 
radiation era evolution. }
\label{fig:amplitude-first-mode}
\end{figure}

We have performed these simulations in a very large lattice of $8192^2$ points.
This allows us to run for a considerably large dynamic range and obtain at the
end of each realization a number of vortices with small enough velocity from which
we can faithfully obtain their level of excitation. Two such examples are given 
in Fig. \ref{fig:amplitude-first-mode} where we display the amplitude of the modes as 
a function of time after the transition to Minkowski space. We have performed $10$ large 
cosmological realizations, and at the end of the simulation we have on average $10\pm4$ vortices,  
and in total 100 vortices. Out of those, 65  were moving slow enough for us to reliably obtain the amplitude 
of the internal modes. As mentioned earlier, we restrict our analysis to the first two bound modes. The average level of the amplitude for each of the bound modes is estimated to be 
\beq
\langle\hat{A}_{1}\rangle \approx 0.049\pm0.036\,\,\,\,\,\,\,,\,\,\,\,\,\,
\langle\hat{A}_{2}\rangle \approx 0.053\pm0.037~.
\label{amplitude-in-scaling}
\eeq

Following the same type of calculation we did earlier for the phase transition we can infer that the vortices are on 
average $0.03\%$ (first mode) and $0.04\%$ (second mode) more energetic due to these
bound states at the end of our simulations. One can also compare the amount of extra energy in the vortex
by considering only the core mass of the vortex. In this case the appropriate percentages are $0.25\%$ and
$0.36\%$ respectively. This is clearly quite a small amount of energy so it is hard to see how this
can change much of the dynamics of these vortices at this point.\\

These results indicate that bound states seem to survive for the duration of our cosmological simulations. Part of this
excitation comes from the initial conditions and part from its subsequent interaction with the radiation and
the other vortices. It is interesting to note that the amplitudes of the modes seem to be significantly lower 
at the end of our cosmological simulations than what we found in our experiments in previous sections 
on the vortex formation during phase transitions. As we mentioned earlier, it would be interesting to 
investigate the dependence of these results with the procedure we use to set the initial conditions,
the expansion history, etc. However, given the low level of excitation of vortices 
in our simulation, it is not clear to us whether these bound states could play a significant role
in this type of models.

\section{Conclusions}

In this paper we have investigated the spectrum of perturbations around $(2+1)$ dimensional global vortices. 
We have found that vortices of unit winding number have localized states whose
oscillating frequency is below the mass of the radial excitation of the field in the vacuum.
This means that they will only radiate at the non-linear level. Using similar techniques
to the ones developed in \cite{Manton:1996ex,Blanco-Pillado:2020smt} we were able to compute the decay rate of
 the first two bound state modes analytically. These calculations show that these modes radiate
most of their energy in the form of  massive scalar radiation at double its oscillating frequency.
Furthermore, the typical time scale for their decay is much longer than the period
of oscillation or the only other scale in the problem, the width of the
vortex. Numerical simulations of the excited vortex in a field theory lattice, using the full 
nonlinear equations of motion, show a perfect agreement with these analytic
predictions.

All this means that an excited vortex can store in these internal modes a large
percentage of its core mass for a very long time. It is therefore interesting to
evaluate this possibility quantitatively since it could in principle affect the dynamics
of the vortices in a realistic setting.

We begin our investigations of these effects by demonstrating several different ways that a vortex can get excited
dynamically. We first irradiate the vortex solution with wave packets and compute
the amplitude of the different modes after the scattering between the wave and the vortex has occurred. The
results show that this excitation happens for either massless and massive 
radiation. Furthermore, the long range interaction of vortices due to their coupling
to the massless Goldstone mode induces forces between them. The presence of
acceleration also leads to excitations. However, the amplitude of the excited state
can also decrease due to the interaction with radiation, so it is very hard
to predict the level of excitations that these vortices could reach in a dynamical
background. In order to do this, we have designed a few numerical experiments that
can be considered representatives of the different possible realistic situations where
we might be interested to evaluate this process.

We first look at the level of excitation of vortices after their formation in a phase transition.
In order to do that we perform a simulation where the potential for the field changes
abruptly in time leading to the formation of several vortices and anti-vortices. We found that
vortices are indeed excited but the amount of extra energy is quite low. In fact,
the total energy stored in the core of the vortex is not increased
in more than a few percent.

We also obtained the level of excitation of these modes when the vortex is in
contact with a thermal bath. The results also indicate that the transferring of
energy from the thermal background to the internal mode is not very efficient
so the amount of energy is again below the $1\%$ percentage level for all temperatures.
This quantity saturates as one increases the temperature and its value is
in fact very similar to the one obtained in their formation at  phase transitions.

Finally, we simulated vortices in an expanding background. In this case,
their dynamics leads some vortices and antivortices to annihilate and
disappear leaving behind massive radiation that excites other vortices.
Moreover, the interactions between vortices lead them to gain some
extra internal energy as well. However, the result at the end of the simulation 
is again that vortices do not seem to acquire a large amount of extra energy in the
form of internal perturbations. In fact, the
level of excitation is so small that is hard to see how this effect could become
relevant for these models.

The global vortices we described here can be seen as the cross section
of an infinite straight global string, so the perturbation modes we have
been discussing are indeed present on the $(3+1)$ dimensional string as well.
One can decompose the general perturbations for strings in modes of different
wavelengths along the $z$ direction. Doing that we will arrive at a similar
equation to the one we found in this paper for the transverse
part of the mode functions. In particular, we will also find bound states with the same profile along the perpendicular 
directions to the string to the ones described here for modes with wavelengths along the 
z direction large compared to the thickness of the string core.
This means that part of the conclusions that we have arrived here can be extended
to the $(3+1)$ dimensional case. Specifically, the typical time scales
of the perturbations will be similar to the case studied here. However, it is difficult
to estimate whether the low level of excitation seen here in the vortex
case can be extrapolated to the more complicated string network. One
could be tempted to argue that vortices here behave similarly to what
long strings on the  $(3+1)$ network do, and in some regards this is
true. On the other hand, we think that there are important processes that are
missed by our simulations in $(2+1)-$dimensions. In particular,
strings have light degrees of freedom living on their worldsheet,
the Goldstone modes of the position of the string. These wiggle modes could
interact and lead to the excitation of the internal modes. Furthermore, 
strings can intercommute, exchanging partners and producing new
connections between different strings. These reconnections lead to 
kink formation that travel along the string and could excite these
internal modes \footnote{This is in fact what was suggested in \cite{Saurabh:2020pqe}. It would be 
very interesting to see whether the energy in that case decays following the time
scale we obtained in this paper.}.  These processes are not present in our $2+1$ simulations.
It is therefore interesting to investigate whether or not these internal states
get more populated in that case. Some work in this direction is
already underway.

Finally, we would also like to comment on the possible applications of
our methods for condensed matter vortices. The dynamical results we report
here only applied to the relativistic models for vortices that we presented
in Section (II). However, the calculation needed to identify bound states of 
global vortices may be useful in the computation of perturbations of the
vortices in condensed matter theories such as the Gross-Pitaevskii 
equation for superfluids. We hope to be able to explore this aspect
of our work in the near future.

\section{Acknowledgements}

We are grateful to Mark Hindmarsh, Asier Lopez-Eiguren, Ken D. Olum, Tomasz Romańczukiewicz, 
Mikel A. Urkiola and Tanmay Vachaspati for useful discussions.  This work is supported in part by 
the Spanish Ministry MCIU/AEI/FEDER grant (PGC2018-094626-B-C21), the Basque Government 
grant (IT-979-16) and the Basque Foundation for Science (IKERBASQUE). The numerical work 
carried out in this paper has been possible thanks to the computing infrastructure of the ARINA 
cluster at the University of the Basque Country (UPV/EHU).

\appendix

\section{Perturbations around the global vortex}
\label{appendix-perturbations}

As we showed in the main part of the text, the solution describing a global vortex is given
by
\beq
\phi_v({\bf x}) = f(r) e^{i \theta}~.
\eeq
Any generic perturbation around this background solution can be split into its modification 
of the modulus of the field and the variation of the phase.  Such separation makes sense 
especially when we look at the solution asymptotically. There, these degrees of freedom 
decouple and one can identify each of these fluctuations as part of the massive or the massless 
sector in the vacuum. Here we will follow this procedure\footnote{This approach is analogous to 
the analysis of the spectrum of perturbations done in \cite{Goodband:1995rt} for global strings.
}. 

We can also classify the possible perturbations as a function of their azimuthal
dependence. Let us start with the cylindrically symmetric perturbations. In particular, let us
consider a fluctuation of the form
\beq
\phi({\bf x},t) = \left(f(r) + s^r_0(r) \cos\left(\omega t\right) \right) e^{i \theta}\,,
\eeq
where we are taking the function $s^r_0(r)$ to be a real function of the
radial coordinate $r$. This perturbation is nothing more than a fluctuation
on the profile function for the massive part of the vortex solution. The appropriate linearized
equation of motion for this function is given by
\beq
-\nabla_r^2 s_0^{r}+\frac{1}{r^2}s_0^{r}+\frac{1}{2}\left(3f^2-1\right)s_0^{r}=\omega^2s_0^{r}.
\eeq
One can gain some insight about the possible eigenvalues of this equation by drawing from our intuition about the
analogous Schr\"odinger type equation with a potential $U(r)  = \frac{1}{2}\left(3f^2-1\right) + \frac{1}{r^2}$. 
Note that the potential well goes asymptotically to $U (r\rightarrow \infty ) = 1$, so we can expect that the spectrum has a set of bound states with
$\omega^2 < 1$, and a continuum of states beyond this point. This continuum will match asymptotically the 
vacuum solutions obtained by setting $f=1$ in the previous equations, which in our case reduce to the Bessel
functions $J_1(k r)$ where $\omega^2 = k^2 +1$. These are nothing more than the massive states associated
with the radial mode of the field  written in a cylindrically symmetric way.

Using the numerical solution obtained for $f(r)$ one can construct these effective potentials and
look for bound states numerically. Due to the asymptotic behavior of the potential, it is possible to 
show that there are infinitely many bound modes below the threshold. We have computed numerically the first 
two such bound states with eigenvalues\bea
\omega_1^2 = 0.8133,\, \\
\omega_2^2 = 0.9979~.
\eea
 Bound modes above with $\omega>\omega_2$ are very delocalized with respect to the vortex 
core\footnote{See the details that characterize the infinite set of states for a potential
decaying as $U(r) \approx 1 - \frac{2}{r^2}$ in \cite{MF}, vol II, page 1665.}
 This property makes them very hard to study numerically, specially in 2d simulations. We have therefore restricted 
our numerical analysis to the first two modes as representatives of the linear spectra and extrapolated some general properties.
We show in Fig. {\ref{fig:first bound}  the profiles of both mode 
functions.  We denote these mode functions as $s^{(1)}(r)$ and $s^{(2)}(r)$.

Let us now turn to perturbations associated with a small variation of the phase of the field. Such fluctuations
 would lead to an excited field configuration of the form
\beq
\phi({\bf x},t) = \left(f(r) + i  s^i_0(r) \cos\left(\omega' t\right) \right) e^{i \theta}\,,
\eeq
where, in this case, the equation of motion for the perturbation is 
\beq
-\nabla_r^2 s_0^{i}+\frac{1}{r^2}s_0^{i}+\frac{1}{2}\left(f^2-1\right)s_0^{i}=\omega'^2s_0^{i} \, .
\eeq
 
The equation for $s^i_0(r)$ does not have any bound state \footnote{Note that the asymptotic decay of the potential
is faster than the case of the massive modes described earlier that is why there is no infinite set of bound states.}. Furthermore, its continuum spectrum
starts at $\omega '= 0$ and the asymptotic states represent the excitations associated with the massless Goldstone
mode in the vacuum. The solutions are again written in terms of Bessel functions with a different relation of
$\omega'^2 = k^2$. Finally, there is also a zero mode solution of the form  $s^i_0(r) = f(r)$, which just signals the invariance of our
theory with respect to rigid phase rotations of the full background solution.

 We can also consider perturbations with an azimuthal angular dependence. This corresponds
 to field configurations of the form
 \beq
\phi({\bf x},t) = \left(f(r) + \left[ s^r_m(r) \cos(m \theta) + i s^i_m(r) \sin(m \theta) \right] \cos\left(\omega t\right) \right) e^{i \theta}.
\eeq
 
 In this case, the real and imaginary parts of the general fluctuation do not decouple so one is led to an eigenvalue
 problem for a system of differential equations of the form
\bea
&& -\nabla_r^2 s_m^r + \frac{s_m^r}{r^2}\left(m^2+1\right)+\frac{1}{2}\left(3 f^2(r)-1\right)s_m^r+\frac{2m}{r^2}s_m^i=\omega^2 s_m^r,\\
&& -\nabla_r^2 s_m^i + \frac{s_m^i}{r^2}\left(m^2+1\right)+\frac{1}{2}\left( f^2(r)-1\right)s_m^i+\frac{2m}{r^2}s_m^r=\omega^2 s_m^i.
\eea

We have scanned these equations for solutions with eigenvalues $\omega< 1$, which would
denote the presence of a bound state for this system. Our numerical investigations show that there are no
further bound state solutions for $m<10$ apart from the $m=1$ zero modes associated with the rigid translations of
the vortex on the $2d$ plane. Although we do not have a definite proof we conjecture that indeed there 
are no bound states in this system for larger values of $m$ either.

Furthermore, strings with higher winding ($N>1$) have a number of negative eigenvalues ($m>1$) corresponding to the splitting 
 of the string into lower-charge strings (see \cite{Goodband:1995rt}).

\section{Asymptotic form of the radiation}
\label{radiation-field}

As we described in the main part of the text, one can obtain an equation for the radiation
modes ($\eta(r,t)$) sourced by the oscillating bound state ($s(r)$) of the form
\bea\label{eigen-rad-re-app}
&&\pa_t^2 \eta(r,t)-\pa_r^2  \eta(r,t)-\frac{\pa_r  \eta(r,t)}{r}+\frac{\eta(r,t)}{r^2}+\frac{1}{2}\left(3 f^2(r)-1\right)\eta(r,t)=\\
&&\frac{3}{2}A^2(t)\left(\alpha s(r)-s(r)^2f(r)\right)\nonumber.
\eea
Taking the radiation modes to be described by
\be
\eta(r,t)=g(r) e^{- 2 i \omega t}~,
\ee
and separating the fast and slow variation of the amplitude of the source using the parametrization
\beq
A(t) = \hat A(t) \cos\left( \omega t\right)\,,
\eeq
we arrive to the equation
\bea\label{eigen-rad-re1-app}
&&-g''(r)-\frac{g'(r)}{r}+\frac{g(r)}{r^2}+\frac{1}{2}\left(3 f^2(r)-1\right)g(r)-(2\omega)^2 g(r)=\label{eq-rad-app}\\
&&\frac{3}{4}A_0^2\left(\alpha s(r)-s(r)^2f(r)\right)\nonumber,
\eea
where $A_0=\hat A(0)$.
We will solve this equation by the Green's function method similarly to what was done in the $1+1$ analogous problem
in \cite{Manton:1996ex}. In order to do that we first compute the asymptotic form of the outgoing solutions for the 
homogeneous equation (\ref{eq-rad-app}) namely,
\bea\label{homoge1-app}
y_1(r)&=&\frac{D}{\sqrt{r}}\cos\left(r\sqrt{4\omega^2-1}+\hat \zeta\right)~,\label{homoge1} \\
y_2(r)&=&J_1(r\sqrt{4\omega^2-1})-i Y_1(r\sqrt{4\omega^2-1})=H_1^{(2)}(r\sqrt{4\omega^2-1})\label{homoge2}~,
\eea
where $H_1^{(2)}(r)$ is Hankel function associated with a 
particular combination of the Bessel functions $J_1(r)$ and $Y_1(r)$, and $D$ and $\hat \zeta$ are the amplitude and phase determined by the initial conditions.  

Using the method of variation of parameters we can obtain a particular solution of (\ref{eigen-rad-re1-app}). This solution 
can be written in the form
\be
g(r)=-z_2(r)\int_0^r \frac{F(r')z_1(r')}{W(r')} dr' -z_1(r)\int_r^\infty \frac{F(r')z_2(r')}{W(r')} dr'~,
\ee 
where $F(r)$ is the non-homogenous part in (\ref{eq-rad-app}), $W(r)$ is the Wronskian related to (\ref{homoge1}) 
and (\ref{homoge2}) and $z_{1}(r)$ is an exact solution of the homogeneous equation whose asymptotic form is given by (\ref{homoge1}) with $z_{1}(0)=0$. We choose $z_2(r)$ as a complex solution which asymptotically is given by (\ref{homoge2}). Since we are interested 
in the outgoing radiation, we take the solution for large $r$ to be approximated by
\be\label{etaapp}
g(r)\approx-y_2(r)\int_0^\infty \frac{F(r')z_1(r')}{W(r')}dr'~.
\ee

On the other hand, the Wronskian of two solutions of the homogenous part of  (\ref{eq-rad-app}) can be computed exactly to give
\be
W(r)=C/r.
\ee
Since we know that $y_1(r)$ and $y_2(r)$ are exact solutions for large $r$, we can determine the coefficient $C$ imposing 
that the asymptotic behavior of the complete solutions is that of (\ref{homoge1}) and (\ref{homoge2}). This gives
\be
W(r)=-\frac{0.63+i\, 0.18}{ r}.
\ee

Finally, we can compute the integral in (\ref{etaapp}) using a numerical solution for $z_1(r)$ that verifies the asymptotic 
behavior (\ref{homoge1}), while for $y_2(r)$ we can use the asymptotic expansion for the Hankel function. We get from (\ref{etaapp})
\be\label{rad1-app}
g(r)=\sqrt{\frac{2}{\pi r \sqrt{4\omega^2-1}}}e^{i r\sqrt{4\omega^2-1}+i\zeta} \vert \gamma\vert=\frac{\Delta(\gamma,\omega)}{\sqrt{r}}e^{i r\sqrt{4\omega^2-1}+i\zeta},
\ee 
where $\gamma$ is the numerical factor given by the integral in (\ref{etaapp}), and $\Delta(\gamma,\omega)$ can be read from
(\ref{rad1-app})
\beq 
\Delta(\gamma,\omega) = \sqrt{\frac{2}{\pi  \sqrt{4\omega^2-1}}}\vert \gamma\vert\,.
\eeq
The values of these coefficients for our two modes are  $\Delta_1=0.0256$ and $\Delta_2 =0.00028$.
We have also included a phase denoted by $\zeta$ which is not relevant in the rest of the calculation.

Putting all things together we arrive to the result of the asymptotic form of the radiation field induced
by the oscillating internal modes:
\be
\eta(r,t)\approx  \hat A^2(t) \text{Re}
\left[\frac{\Delta(\gamma,\omega)}{\sqrt{r}} e^{-i 2\omega t+i r\sqrt{4\omega^2-1}+i\zeta}\right]=  \hat A^2(t) \frac{\Delta(\gamma,\omega)}{\sqrt{r}}\cos\left( 2\omega t- r\sqrt{4\omega^2-1}-\zeta\right) . \\
\ee 
\\
From the $0r$ component of the stress-energy tensor we can determine the energy flux 
in the radial direction, using expression (\ref{energy-flux}). The average energy flux computed over a period (which adds a factor $1/2$) is given by 
\be
\langle T_{0 r} \rangle = -\frac{4\omega}{\pi r}\vert \gamma\vert^2 \hat A^4\equiv-\frac{b}{r}\hat{A}^{4}
\ee
for large $r$. Finally, the power radiated to infinity is
\be
\dot{E}=\int_{0}^{2\pi} \langle T_{0 r} \rangle \,r\,d\theta=-2\pi b\hat{A}^{4}.
\ee

\section{Details of the numerical simulations}
\label{appendix-numerics}

As it has been mentioned in the main text, we solve the equations of motion for the field in a $1+1$ or a $2+1$ dimensional lattice depending on the effect we want to study. On the one hand, when the problem can be simulated using cylindrical symmetry,  the $1+1$ dimensional lattice can be used. This case allows for a much bigger number of lattice points, and thus, bigger accuracy and dynamical range. However, this approach pins the core of the string to a point, and does not allow for motion of the vortex. 

On the other hand,  $2+1$ dimensional lattices can be used to solve all problems. This allows for situations with no symmetry, it allows for the core to move, and we can have more than one vortex in the simulation. The price to pay is a heavier numerical budget, and thus the accuracy and dynamical range is lower than the 1+1 dimensional case.

Both types of simulations present some computational challenges. Firstly, regardless of the number of spatial dimensions, we will need a large box for the second mode to fit in. Secondly, for the $2+1$ simulations, we will implement the evolution in an expanding background. In this case, the fact that the vortices have fixed physical size implies that they shrink in comoving coordinates, so we may run out of resolution in their cores if we evolve the system for long periods of time. And even if we do this for shorter time scales, this may still be a problem if the expansion rate of the universe is big enough. Therefore, a large number of lattice points is needed to obtain an accurate description of the dynamics of the system. In order to face these issues, we wrote a parallel code using  \emph{message passing interface} (MPI). This allows us to run the much larger lattice simulations we need in this work.

Another key feature of our simulations is the employment of absorbing boundary conditions (described below), which allow us to alleviate the problem of the interaction of the vortices with radiation that is reflected at the boundaries of the simulation box.

\subsubsection{$1+1$ simulations}

This type of simulations has been used when the problem presented a cylindrical  symmetry.  Then, the solution of the equations of motion at any time is of the form $\phi\left(r,\theta,t\right)=F\left(r,t\right)e^{i\theta}$,  which reduces to  the following radial equations of motion for  the real and imaginary parts  of the field, $\phi=\phi_1+i \phi_2$:
\begin{equation}
\ddot{\phi}_{1,2}=\frac{\partial^{2}\phi_{1,2}}{\partial r^{2}}+\frac{1}{r}\frac{\partial\phi_{1,2}}{\partial r}-\frac{\phi_{1,2}}{r^{2}}-\frac{1}{2}\phi_{1,2}\left(\phi_{1}^{2}+\phi_{2}^{2}-1\right).
\label{eq:radial eq one d}
\end{equation}

We have solved these equations using standard discretization techniques, using MPI.  The boundary conditions applied were $\phi(r=0)=0$ and absorbing boundary conditions at $r=L$, where $L$ is the size of the simulation box. Absorbing boundary conditions are the best suited for the problem at hand, because  we do not wish that the radiation that the vortex emits  bounces off the simulation edges back to the vortex.

A boundary condition of the form 
\begin{equation}
\frac{\partial\phi_{1,2}}{\partial t}+\frac{\partial\phi_{1,2}}{\partial r}\,\,\,\,\bigg\rvert_{r=L}=0\,\,,
\label{eq:abc cylindrical symmetry}
\end{equation}
would absorb an outgoing cylindrical wave at an asymptotically far boundary. Such a wave would be given by
\begin{equation}
\xi\left(r,t\right)\propto\frac{1}{\sqrt{r}}\cos\left(\beta t-kr+\delta\right).
\label{eq:cylindrical wave}
\end{equation}
One can understand why Eq.~(\ref{eq:abc cylindrical symmetry}) is a good absorbing boundary condition by noticing that 
$\phi_{1,2}=1+\xi$ is its approximate solution for $\beta=k$. This last condition implies that the absorbing boundary condition works better for modes with $k\gg m=1$.

One can refine this condition by tailoring the equation to be satisfied at the boundary for a cylindrical monochromatic wave with known angular frequency $\beta$, namely:\\} 
\begin{equation}
\frac{\partial\phi_{1,2}}{\partial t}+\frac{\beta}{\sqrt{\beta^{2}-1}}\frac{\partial\phi_{1,2}}{\partial r}+\frac{\beta}{2\sqrt{\beta^{2}-1}}\frac{\phi_{1,2}-1}{r}\,\,\,\,\bigg\rvert_{r=L}=0\,.
\label{eq:abc refined}
\end{equation}
\\
We have checked in our simulations that both Eq. (\ref{eq:abc cylindrical symmetry}) and Eq. (\ref{eq:abc refined}) yield similar results; both choices 
seem to be equally efficient at absorbing outgoing radiation.

There have been several instances in which the 1+1 dimensional approach has been useful, mainly in the determination of the decay time scale of the bound modes and the study of their excitation by illuminating the vortex with radiation.

In the study of the decay of the bound modes (see Section (\ref{exci})), we have solved Eq. (\ref{eq:radial eq one d}) for an initial condition of the type given by Eq. (\ref{eq:vortex plus shape}). For the case where the initial condition only had the first mode, i.e., $A_1=0.2$ and $A_2=0$, we chose $L=40$ and $\Delta r=0.004$ (with $\Delta t=0.001$). For the case  where only the second mode was initially excited, we used $L=200$ and $\Delta r=0.05$ (with $\Delta t=0.01$). The typical number of time steps needed for this study was huge: $\sim 10^7$ for the first mode and $\sim 10^9$ for the second one. 

In the experiments consisting of illuminating the vortex (studied in Section (\ref{ilumin1}) and Section (\ref{ilumin2})), we used $L=60$, $\Delta r=0.003$ and $\Delta t=0.001$. In these cases, we usually ran the simulations for $2\times10^{5}$ time steps.

\subsubsection{$2+1$ simulations}

For these simulations we do not use any symmetry, so we solve numerically the equations of motion for the real and imaginary parts of the complex field $\phi=\phi_{1}+i\phi_{2}$ in an expanding background.

The equations for cosmic time read
\beq
\ddot \phi_{1,2} + 2H \dot \phi_{1,2} - \frac{1}{a^2}\nabla^{2}\phi_{1,2} + \frac{1}{2} \phi_{1,2} \left(\phi_{1}^{2}+\phi_{2}^{2} - 1\right)= 0~,
\label{eq of motion}
\eeq
where dots denote partial derivatives with respect to cosmic time and $\nabla^{2}=\partial_{x}^{2}+\partial_{y}^{2}$, with $x$ and $y$ being comoving coordinates; $a$ is the scale factor, and $H=\dot{a}/a$ is the Hubble rate. 

A flat space version of these equations of motion in $2+1$ dimensions has been used for all the cases studied in the $1+1$ case to make sure that there was not extra dynamics that was lost while imposing cylindrical symmetry. In order to do that, we used the corresponding boundary conditions implemented for the $2+1$ dimensional simulations, which are either the Cartesian version of Eq. (\ref{eq:abc cylindrical symmetry}), 
\begin{equation}
\frac{\partial\phi_{1,2}}{\partial t}+\frac{\partial\phi_{1,2}}{\partial x}\frac{x}{\sqrt{x^{2}+y^{2}}}+\frac{\partial\phi_{1,2}}{\partial y}\frac{y}{\sqrt{x^{2}+y^{2}}}\,\,\,\,\bigg\rvert_{x=\pm L/2, y=\pm L/2}=0\,\,,
\label{eq:abc radial}
\end{equation}
\\
which is more efficient at absorbing modes with radial incidence, or
\begin{equation}
\frac{\partial\phi_{1,2}}{\partial t}\pm\frac{\partial\phi_{1,2}}{\partial x}\,\,\,\,\bigg\rvert_{x=\pm L/2}=0\,\,,\,\,\,\,\,\,\,\,\,\frac{\partial\phi_{1,2}}{\partial t}\pm\frac{\partial\phi_{1,2}}{\partial y}\,\,\,\,\bigg\rvert_{y=\pm L/2}=0
\label{eq:abc normal}
\end{equation}
which is more efficient at absorbing modes with normal incidence.

The $2+1$ version was indispensable for Section (\ref{vavi}), where we simulated a vortex-antivortex interaction, and for Section (\ref{cosmoevol}). 

\begin{figure}[h!]
\includegraphics[width=8cm]{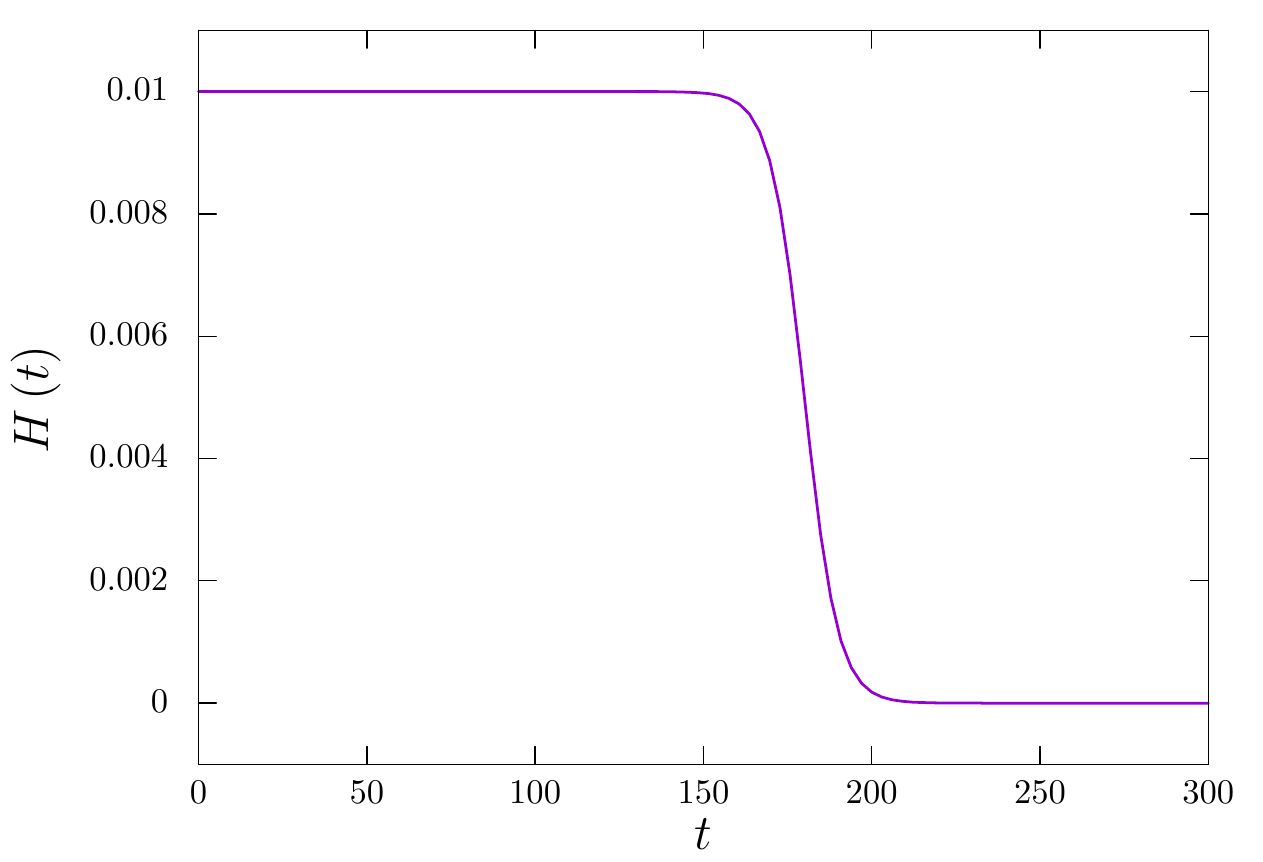}
\includegraphics[width=8cm]{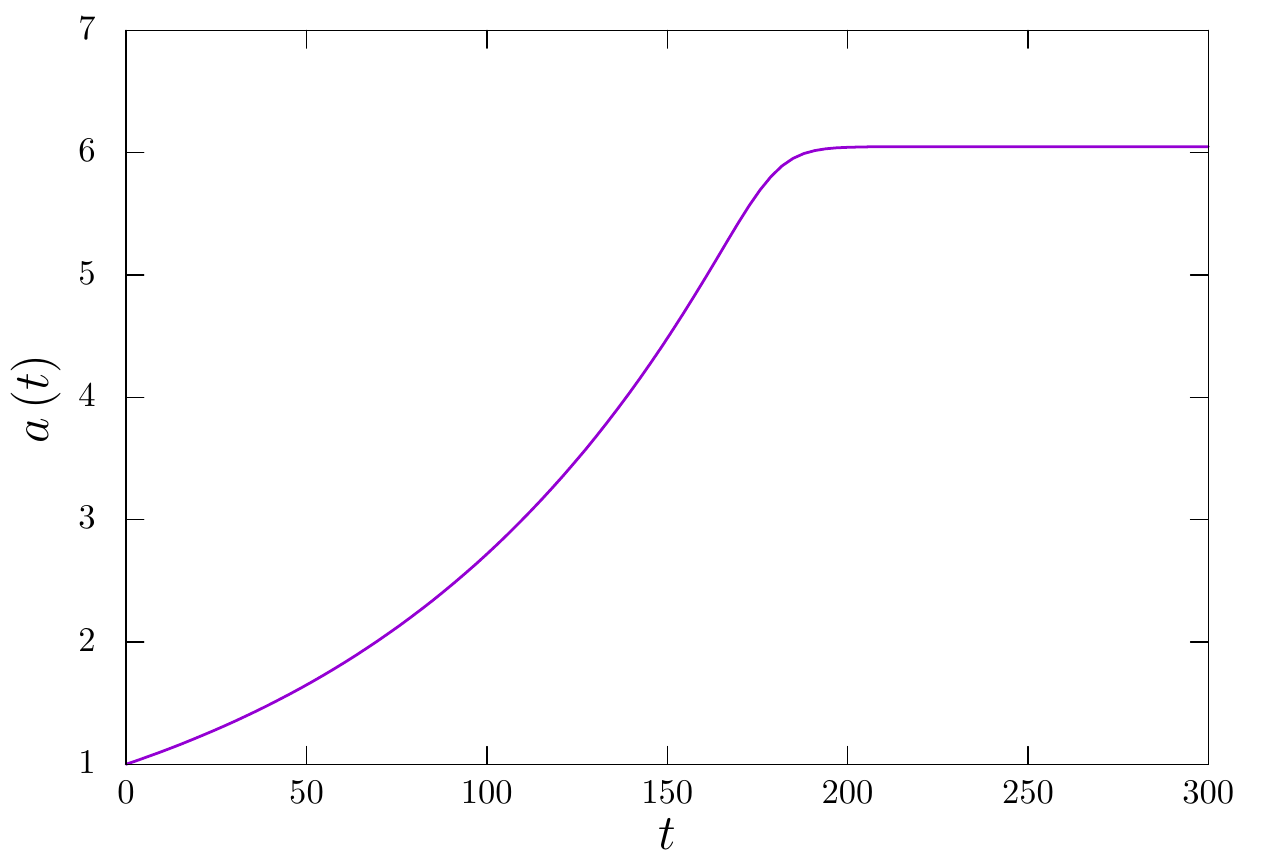}
\caption{Hubble rate (left) and scale factor (right) as a function of cosmic time.}
\label{fig:expansion universe}
\end{figure}

The simulations in Section (\ref{at formation})  also rely on $2+1$ simulations. Their initial conditions  are rather involved and we describe them in the subsequent section.  After the initial condition has been set, both the  simulations have been evolved solving Eq. (\ref{eq of motion}) in an expanding Universe.
For both, we chose a Hubble rate of the form
\begin{equation}
H\left(t\right)=\frac{H_{0}}{2}\left[1-\tanh\left(\frac{t-t_{*}}{\Delta}\right)\right]\,,
\label{eq:hubble rate}
\end{equation}
which corresponds to a scale factor
\begin{equation}
a\left(t\right)=e^{\frac{H_{0}}{2}\,t}\left[\frac{\cosh\left(\frac{t_{*}}{\Delta}\right)}{\cosh\left(\frac{t-t_{*}}{\Delta}\right)}\right]^{\frac{H_{0}\Delta}{2}}\,.
\label{eq:scale factor}
\end{equation}
These functions allow for a smooth transition from de Sitter space to Minkowski space. We plot them in Fig.~\ref{fig:expansion universe}
 for our choice of dimensionless parameters: $H_{0}=0.01$, $t_{*}=180$ and $\Delta=10$.\\

These simulations were typically run for 7500 time steps, and the scale factor has grown by a factor of 6 when we reach the Minkowski stage.

\section{A thermal state on the lattice}
\label{lattice-thermal-state}

In this section we describe how the initial conditions have been set for Section (\ref{at formation}).
{\subsection{Phase transition}\label{t1}}

In order to simulate the phase transition, we will make the potential change abruptly at $t=0$ from a quadratic potential $V_{1}$ to the usual Mexican hat potential $V_{2}$:
\begin{equation}
\label{2pot}
V(\phi_{1},\phi_{2}) =
\begin{cases} 
V_{1}\left(\phi_{1},\phi_{2}\right)=\frac{\lambda \eta^4}{4}+m_{r}^2\left(\phi_{1}^2+\phi_{2}^2\right)&\textrm{for $t<0$}\\\\
V_{2}\left(\phi_{1},\phi_{2}\right)=\frac{\lambda}{4}\left(\phi_{1}^2+\phi_{2}^2 -\eta^2\right) ^2&\textrm{for $t\geq0$}
\end{cases}
\end{equation}
where $m_{r}^{2}=\lambda\eta^{2}$ is chosen to be the squared mass of the Higgs boson of the $V_{2}$ potential theory and $\phi_{1}$ and $\phi_{2}$ respectively denote the real and imaginary part of $\phi$. Note that in the $V_{1}$ theory $m_{r}$ is the mass of the fields $\Phi_{1,2}\equiv\sqrt{2}\phi_{1,2}$. Now we aim to construct a thermal state for $\Phi_{1}$ and $\Phi_{2}$ at the bottom of the $V_{1}$ potential. This will be the initial state for the subsequent evolution under the influence of the $V_{2}$ potential.

Let $\xi_{1,2}$ be the fields representing the thermal fluctuations around $\Phi_{1,2}=0$, and let $\pi_{1,2}$ be their corresponding velocities. As the derivation of the thermal state is completely analogous for both fields $\Phi_{1}$ and $\Phi_{2}$, we will only focus on the former. The fluctuation fields can be expanded as follows:
\begin{equation}
\xi_{1}\left(t=0,x,y\right)=\sum_{m}\sum_{n}\frac{1}{\sqrt{2L_{x}L_{y}\omega_{mn}}}\left[\alpha_{mn}\,e^{i\left(k_{x}^{(m)}x\,+\,k_{y}^{(n)}y\right)}+\alpha_{mn}^{*}\,e^{-i\left(k_{x}^{(m)}x\,+\,k_{y}^{(n)}y\right)}\right],
\label{eq:tic phi phys}
\end{equation}
\begin{equation}
\pi_{1}\left(t=0,x,y\right)=\sum_{m}\sum_{n}\frac{1}{i}\sqrt{\frac{\omega_{mn}}{2L_{x}L_{y}}}\left[\alpha_{mn}\,e^{i\left(k_{x}^{(m)}x\,+\,k_{y}^{(n)}y\right)}-\alpha_{mn}^{*}\,e^{-i\left(k_{x}^{(m)}x\,+\,k_{y}^{(n)}y\right)}\right],
\label{eq:tic pi phys}
\end{equation}
\\
where $k_{x}^{(m)}=2\pi m/L_{x}$, $k_{y}^{(n)}=2\pi n/L_{y}$ and
\begin{equation}
\omega_{mn}=\sqrt{\left[\frac{2\sin{\left(\frac{k_{x}^{(m)}\Delta x}{2}\right)}}{\Delta x}\right]^{2}+\left[\frac{2\sin{\left(\frac{k_{y}^{(n)}\Delta y}{2}\right)}}{\Delta y}\right]^{2}+m_{r}^{2}}\,\,.
\label{eq:omega phys tic}
\end{equation}
In the previous expressions, the integer $m$ runs from $-N_{x}/2+1$ to ${N_{x}/2}$, and $n$ runs from $-N_{y}/2+1$ to ${N_{y}/2}$, where $N_{x}$ and $N_{y}$ are the number of lattice points in each spatial dimension: $N_{x}=L_{x}/\Delta x$, $N_{y}=L_{y}/\Delta y$.

The complex coefficients $\alpha_{mn}$ are given by a Gaussian Random Field with two point function 
\begin{equation}
\langle|\alpha_{mn}|^{2}\rangle=\frac{1}{e^{\omega_{mn}/T}-1}=\frac{1}{2}\left[\coth{\left(\frac{\omega_{mn}}{2T}\right)}-1\right]\,\,,
\label{eq:occupation number tic phys}
\end{equation}
\\
which corresponds to the Bose-Einstein distribution.\\

After introducing the dimensionless variables $\tilde{\xi}_{1}=\xi_{1}/\eta$, $\tilde{L}_{x,y}=m_{r}L_{x,y}$, $\tilde{x}^{\mu}=m_{r}x^{\mu}$, $\Delta\tilde{x}=m_{r}\Delta x$, $\Delta\tilde{y}=m_{r}\Delta y$, $\tilde{k}_{x,y}=k_{x,y}/m_{r}$, $\tilde{\omega}_{mn}=\omega_{mn}/m_{r}$, $\tilde{\alpha}_{mn}=\left(\sqrt{m_{r}}/\eta\right)\,\alpha_{mn}$ and the dimensionless temperature $\Theta=T/\eta^{2}$, the expression given by Eq. (\ref{eq:occupation number tic phys}) can be rewritten as
\begin{equation}
\langle|\tilde{\alpha}_{mn}|^{2}\rangle=\frac{\sqrt{\lambda}}{2\eta}\left[\coth{\left(\frac{\sqrt{\lambda}\,\tilde{\omega}_{mn}}{2\eta\Theta}\right)}-1\right]\,,
\label{eq:average dimensionless}
\end{equation}
which tends to $\Theta/\tilde{\omega}_{mn}$ in the classical limit ($T\gg\omega_{mn}$).\\

Finally, further simple manipulations yield the following expressions for the thermal fluctuations:
\begin{equation}
\tilde{\xi}_{1}\left(\tilde{t}=0,\tilde{x},\tilde{y}\right)=\sum_{m}\sum_{n}\sqrt{\frac{\langle|\tilde{\alpha}_{mn}|^{2}\rangle}{\tilde{L}_{x}\tilde{L}_{y}\tilde{\omega}_{mn}}}r_{mn}\cos\left(\tilde{k}_{x}^{(m)}\tilde{x}\,+\,\tilde{k}_{y}^{(n)}\tilde{y}+\gamma_{mn}\right),
\label{eq:xi dimensionless final}
\end{equation}
\begin{equation}
\tilde{\pi}_{1}\left(\tilde{t}=0,\tilde{x},\tilde{y}\right)=\sum_{m}\sum_{n}\sqrt{\frac{\tilde{\omega}_{mn}\langle|\tilde{\alpha}_{mn}|^{2}\rangle}{\tilde{L}_{x}\tilde{L}_{y}}}r_{mn}\cos\left(\tilde{k}_{x}^{(m)}\tilde{x}\,+\,\tilde{k}_{y}^{(n)}\tilde{y}-\gamma_{mn}\right),
\label{eq:xi dot dimensionless final}
\end{equation}\\
Note that the amplitude of the thermal modes is proportional to $\sqrt{\Theta}$ in the classical limit. Note also that the amplitude of those modes with $\omega_{mn}\gg T$ is suppressed, according to the Bose-Einstein distribution. The condition $\omega_{mn}\gg T$ is equivalent to $\tilde{\omega}_{mn}\gg\left(\eta/\sqrt{\lambda}\right)\Theta\equiv\tilde{\Omega}$.  Therefore, modes with angular frequencies much greater than the cut-off frequency $\tilde{\Omega}$ are suppressed.

The Eq. (\ref{eq:average dimensionless}) can be rewritten in terms of $\tilde{\Omega}$ and $\Theta$:
\begin{equation}
\langle|\tilde{\alpha}_{mn}|^{2}\rangle=\frac{\Theta}{2\tilde{\Omega}}\left[\coth{\left(\frac{\tilde{\omega}_{mn}}{2\tilde{\Omega}}\right)}-1\right]\,.
\label{eq:average dimensionless final}
\end{equation}
Thus, we will only need to specify the values of $\tilde{\Omega}$ and $\Theta$ in order to generate the fluctuations by means of Eq. (\ref{eq:xi dimensionless final}) and Eq. (\ref{eq:xi dot dimensionless final}), once the geometry of the lattice has been determined. Let us set $\Delta\tilde{x}=\Delta\tilde{y}\equiv\Delta\tilde{l}$ and $\tilde{L}_{x}=\tilde{L}_{y}\equiv\tilde{L}$. The number of lattice points is, then, $N=\left(\tilde{L}/\Delta\tilde{l}\right)^{2}$.

In our simulations, we choose $\tilde{\Omega}$ and $\Theta$ in such a way that the energy of the thermal fluctuations, $E_{thermal}$, is a small fraction of the initial background energy, $E_{vacuum}$. Firstly, suppose that we want to work in the classical limit (that is, with no thermal modes suppressed). This means that no mode satisfies $\tilde{\omega}_{mn}\gg\tilde{\Omega}$, so we have to choose a cut-off frequency $\tilde{\Omega}$ which is greater than the maximum angular frequency allowed by the lattice spacing: $\tilde{\Omega}>\tilde\omega_{max}\approx\sqrt{2}\,\pi/\Delta\tilde{l}$. In this case, the energy in thermal fluctuations is given by\footnote{Each quadratic degree of freedom in the discrete Hamiltonian contributes with $T/2$ to the average thermal energy. Since the number of quadratic degrees of freedom is $4N$ ($\xi_{1}$, $\xi_{2}$, $\pi_{1}$ and $\pi_{2}$ at each lattice point), the average thermal energy is $2NT$. In agreement with equipartition, the average kinetic energy is $NT$, and the average gradient$+$potential energy is also $NT$.}
\begin{equation}
E_{thermal}\sim2NT=2N\eta^{2}\Theta,
\label{eq:thermal energy}
\end{equation}
while the initial vacuum energy is
\begin{equation}
E_{vacuum}=L^{2}V_{1}\left(\phi_{1}=0,\phi_{2}=0\right)=L^{2}\frac{\lambda\eta^{4}}{4}=\frac{\eta^{2}\tilde{L}^{2}}{4}\,\,.
\label{eq:vacuum energy}
\end{equation}
\\
Finally, imposing that $E_{thermal}$ is $p\%$ of $E_{vacuum}$, one gets $\Theta\sim\left(p/800\right)\Delta\tilde{l}$.\\\\
Typically, we chose $p=10$ or smaller in our simulations, and we set $\tilde{L}=200$, $\Delta\tilde{l}=0.1$ ($2000\times2000$ points) and $\Delta\tilde{t}=0.04$. Since the spatial extent of the vortex is $\sim10$, we expect that thermal modes with wavelength smaller than this value will not excite the bound states. Therefore, in practice, we only generated modes with wavelengths greater than $5$. This is way less demanding in computational terms than generating modes of all the allowed frequencies. In this case, if we choose $\Theta\sim\Delta\tilde{l}/80$, it is guaranteed that the thermal energy is even smaller than $10\%$ of the background energy.
\\
\subsection{Vortex in a thermal bath}
\label{t2}

We are interested in generating the thermal fluctuations directly on the vacua of the potential ($V_2$ in Eq.~(\ref{2pot})). In order to do this, consider the decomposition of the complex field, $\phi=\left(\Phi_{1}+i\Phi_{2}\right)/\sqrt{2}$. The Lagrangian reads
\begin{equation}
\mathcal{L}=\frac{1}{2}\partial_{\mu}\Phi_{1}\partial^{\mu}\Phi_{1}+\frac{1}{2}\partial_{\mu}\Phi_{2}\partial^{\mu}\Phi_{2}-\frac{\lambda}{4}\left(\frac{\Phi_{1}^{2}+\Phi_{2}^{2}}{2}-\eta^{2}\right)^{2}~.
\label{eq:lagrangian one}
\end{equation}

Let us consider fluctuations of the field $\Phi_{1}$ (we will denote the fluctuations by $\xi_{1}$) around the vacuum $\Phi_{2}/\sqrt{2}=C$, $\Phi_{1}/\sqrt{2}=\sqrt{\eta^{2}-C^{2}}$:
\begin{equation}
\mathcal{L}\approx\frac{1}{2}\partial_{\mu}\xi_{1}\partial^{\mu}\xi_{1}-\frac{\lambda}{2}\left(\eta^{2}-C^{2}\right)\xi_{1}^{2}\,,
\label{eq:lagrangian two}
\end{equation}
so the squared mass of the field would be $m^{2}=\lambda\left(\eta^{2}-C^{2}\right)$. Note that the mass depends on the particular vacuum at which the oscillations take place. This would not allow us to construct the matrix $\omega_{mn}$ as in the previous section.

Consider instead the fields $\varphi$ and $\delta$ related to the radial and angular parts of $\phi$ in the following way:
\begin{equation}
\phi=\frac{\varphi}{\sqrt{2}}\,e^{i\frac{\delta}{\eta}}\,.
\label{eq:field radial angular}
\end{equation}
In terms of these two fields, the Lagrangian reads
\begin{equation}
\mathcal{L}=\frac{1}{2}\partial_{\mu}\varphi\partial^{\mu}\varphi+\frac{\varphi^{2}}{2\eta^{2}}\partial_{\mu}\delta\partial^{\mu}\delta-\frac{\lambda}{4}\left(\frac{\varphi^{2}}{2}-\eta^{2}\right)^{2}\,.
\label{eq:lagrangian three}
\end{equation}
Considering fluctuations about any vacuum, that is, $\varphi=\sqrt{2}\eta+\xi$ and $\delta=\eta\theta+\chi$ (where $\theta$ is the polar coordinate), 
\begin{equation}
\mathcal{L}\approx\frac{1}{2}\partial_{\mu}\xi\partial^{\mu}\xi+\frac{1}{2}\partial_{\mu}\chi\partial^{\mu}\chi-\frac{\lambda}{2}\eta^{2}\xi^{2}\,,
\label{eq:lagrangian four}
\end{equation}
so $m_{r}^{2}=\lambda\eta^{2}$ and $m_{\chi}=0$. Now we can find the thermal fluctuations $\xi$ and $\chi$ following the procedure described in the previous subsection. The only difference now is that one of the fields is massless. These fluctuations are added to the static solution $\phi=\eta f\left(r\right)e^{i\theta}$, so the initial conditions for the real and imaginary parts are
\begin{equation}
\tilde{\phi}_{1}\equiv\frac{\phi_{1}}{\eta}=\left[f\left(r\right)+\tilde{\xi}\right]\cos\left(\theta+\tilde{\chi}\right)\,,
\label{eq:phi1 heat up}
\end{equation}
\begin{equation}
\frac{\partial\tilde{\phi}_{1}}{\partial\tilde{t}}=\tilde{\pi}_{\xi}\cos\left(\theta+\tilde{\chi}\right)-\left[f\left(r\right)+\tilde{\xi}\right]\tilde{\pi}_{\chi}\sin\left(\theta+\tilde{\chi}\right)\,,
\label{eq:pi1 heat up}
\end{equation}
\begin{equation}
\tilde{\phi}_{2}\equiv\frac{\phi_{2}}{\eta}=\left[f\left(r\right)+\tilde{\xi}\right]\sin\left(\theta+\tilde{\chi}\right)\,,
\label{eq:phi2 heat up}
\end{equation}
\begin{equation}
\frac{\partial\tilde{\phi}_{2}}{\partial\tilde{t}}=\tilde{\pi}_{\xi}\sin\left(\theta+\tilde{\chi}\right)+\left[f\left(r\right)+\tilde{\xi}\right]\tilde{\pi}_{\chi}\cos\left(\theta+\tilde{\chi}\right)\,.
\label{eq:pi2 heat up}
\end{equation}
\\
We also chose $\tilde{L}=200$, $\Delta\tilde{l}=0.1$ and $\Delta\tilde{t}=0.04$ for these simulations.
\\
\section{Simulating the cosmological evolution of the vortex network }
\label{simcos}

In this section we describe how the simulations for Section (\ref{cosmoevol}) were set up and performed. The main equation of motion to be solved is Eq. (\ref{eq of motion}), but in this situation we have used 
a conformal time, $\tau$, namely,
\beq
 \phi''_{1,2} + 2H \phi'_{1,2} - \nabla^{2}\phi_{1,2} + \frac{1}{2} a^2 \phi_{1,2} \left(\phi_{1}^{2}+\phi_{2}^{2} - 1\right)= 0~,
\eeq
where the primes denote derivatives with respect to conformal time.

We have discretized these equations using standard techniques (see for example \cite{Bevis:2006mj}), using the staggered leapfrog method for the time discretization.

Since we are using comoving coordinates, as the simulation evolves, the vortices shrink (comovingly). Thus, we have to make sure that we resolve the cores for the duration of our simulation. In order to do so, we have simulated  the vortex network  following one of the simulation techniques found in the literature (see for example~\cite{Bevis:2006mj,Hindmarsh:2019csc,Hindmarsh:2021vih}). This procedure consists of several steps (we refer to the interested reader to the above citation for details).

The initial conditions we have used have been those of having the scalar field in the vacuum manifold, with no velocity, but with random orientation of its phase. This configuration has a very high energy, and in order to approach a scaling regime, a period of diffusion evolution is used. Then, we employ a period in which the core of the defects grows artificially until it reaches the desired width (coregrowth period).
It is only then that the proper radiation domination period of the simulation starts. Since the measurements of the amplitudes of the internal modes only make sense in Minkowski space, we smoothly transition from radiation to Minkowski, and then read the corresponding amplitudes.

In this work we did not want to exhaustively study this process, or this procedure;  we just wanted to  analyse whether the outcoming vortices would be excited, and to what approximate level. Thus, we have only chosen one appropriate set of parameters, and have simulated the system 10 times  with different initial random conditions, obtaining 100 vortices (on average $10\pm4$). 

The parameters we have used for our simulations are the following: the simulation box was a square of  size $N=8192$ per dimension, and the comoving lattice spacing was $\Delta x=0.2$. The time resolution was $\Delta \tau=0.04$ (during diffusion $\Delta \tau=0.008$).  This choice of parameters ensures that at the end of the simulation, where the vortices have shrunk due to the comoving coordinates and the expansion of the universe, we resolved the core of the vortex with  enough precision to obtain the amplitude of the internal modes successfully.

We have used a period of 100 time units in diffusion, and  the coregrowth period was from $\tau=10$ until $\tau=65$. The radiation domination simulation ran until $\tau=785$ where a smooth transition to Minkowski happened until $\tau=820$, where the amplitudes were read for a further 50 time units. We run for a total dynamic range of $a(\tau_f)/a(\tau_i) \sim 10$. With those numbers we stop the simulation at roughly half light-crossing time, thus avoiding possible finite volume effects that could arise because of the   use periodic boundary conditions. \\

\bibliography{globalstringexcitations.bib}

\end{document}